\documentclass[prd,superscriptaddress,nofootinbib,preprintnumbers,amsmath,amssymb,10pt,twocolumn]{revtex4-1}
\usepackage{amsmath} 
\usepackage{amssymb}
\usepackage{hyperref}
\usepackage{graphicx}
\usepackage{epsfig}
\usepackage{graphics}
\usepackage{slashed}	
\usepackage{color}
\usepackage[lofdepth,lotdepth,caption=false]{subfig}
\usepackage{array}

\usepackage[T1]{fontenc}
\usepackage{textcomp}

\usepackage{array}
\usepackage{graphicx}
\usepackage{mathrsfs}
\usepackage{multirow}
\usepackage[T1]{fontenc}
\usepackage{textcomp}
\usepackage[T1]{fontenc}
\usepackage{textcomp}
\newcommand{\bra}[1]{\left\langle #1 \right|}
\newcommand{\ket}[1]{\left|#1\right\rangle}

\newcommand{\bfq}{ {\bf q} }
\newcommand{\bfe}{ {\bf e} }
\newcommand{\bfk}{ {\bf k} }
\newcommand{\bfu}{ {\bf u} }
\newcommand{\bfv}{ {\bf v} }
\newcommand{\bfl}{ {\bf \ell} }

\newcommand{\deriv}[2]{ \frac{ d #1 }{ d #2 }  }

\definecolor{SeaBlue}{rgb}{0.2,0.2,0.6}
\definecolor{Maroon}{rgb}{0.6,0.2,0.2}
\definecolor{cerulean}{rgb}{0., 0.52,0.65}

\definecolor{connor_green}{rgb}{0.29, 0.6,0.2}

\begin{document}

\title{Daily modulation of low-energy nuclear recoils from sub-GeV dark matter}
\author{Connor Stratman}
\author{Tongyan Lin}
\affiliation{Department of Physics, University of California, San Diego, CA 92093, USA }
\date{\today}

\begin{abstract} \noindent At sufficiently low nuclear recoil energy, the scattering of dark matter (DM) in crystals gives rise to single phonon and multiphonon excitations. In anisotropic crystals, the scattering rate into phonons modulates over each sidereal day as the crystal rotates with respect to the DM wind. This gives a potential avenue for directional detection of DM. The daily modulation for single phonons has previously been calculated. Here we calculate the daily modulation for multiphonon excitations from DM in the mass range 1 MeV--1 GeV.  We generalize previous multiphonon calculations, which made an isotropic approximation, and implement results in the \texttt{DarkELF} package. We find daily modulation rates up to 1--10 percent for an Al$_2$O$_3$ target  and DM mass below 30 MeV, depending on the recoil energies probed. We obtain similar results for SiC, while modulation in Si, GaAs and SiO$_2$ is negligible.
\end{abstract}

\maketitle

\section{Introduction}

Due to the motion of the Earth in the Milky Way, the dark matter has a preferred direction in our reference frame. This dark matter ``wind'' would appear to originate from the direction of Cygnus. The goal of directional detection is to observe laboratory signatures of dark matter (DM) scattering which reveal this preferred direction~\cite{Spergel:1987kx,Vahsen:2021gnb}. Relative to the fixed lab frame, the direction of the wind also rotates over a sidereal day. This allows for DM signals to be distinguished from effects that vary with solar day, such as solar neutrino backgrounds~\cite{OHare:2015utx}. Directional detection would thus provide a powerful new handle on the origin of any potential DM signal, as well as paving the way towards elucidating the DM distribution itself~\cite{Mayet:2016zxu,OHare:2019qxc}.

For dark matter above $\sim 10$ GeV producing nuclear recoils, the scattering process itself is isotropic, but the distribution of directions for the nuclear recoil is correlated with the DM wind. Observing the direction of the nuclear recoil requires measuring its track. This has led to much development of gas-phase time projection chambers where the track is sufficiently long~\cite{Ahlen:2010ub, Tao:2020muh, DRIFT:2016utn, Vahsen:2020pzb, Shimada:2023vky}. It has also been proposed to use nuclear emulsion films for tracking~\cite{NEWSdm:2017efa}, as well as to measure tracks of tens of nanometers in solid state detectors with quantum sensing techniques \cite{Rajendran:2017ynw,Marshall:2020azl}. 

For scattering of sub-GeV dark matter off nuclei, the recoil energy will be lower and their tracks even more challenging to observe. However, alternative approaches to directional detection exist, taking advantage of anisotropies in the scattering process itself. For sufficiently low energy recoils, nuclei are no longer accurately modeled as free targets and the effects of the target material must be included. For an anisotropic solid state target, this leads to DM scattering rates which depend on the direction of the incident DM relative to the crystal. As the Earth rotates, the result is a daily modulation of the scattering rate over sidereal day.

Several origins for daily modulation in sub-GeV nuclear recoils have been explored. One proposal relies on the idea that energy thresholds for defect production vary along different directions in the crystal~\cite{Budnik:2017sbu,Sassi:2021umf,Dinmohammadi:2023amy}, so that the rate for DM to produce certain defects should vary with time. This effect is relevant for recoils of $\sim$10 eV energy, sufficient to dislocate a nucleus from its lattice site, and requires a new approach to direct detection involving imaging the defects. 

At even lower energy, DM scattering does not displace nuclei, but instead the energy is deposited into phonons, the collective excitations of atoms. Another class of proposals is to search for single phonon excitations from DM-nucleus interactions~\cite{Knapen:2017ekk, Griffin:2018bjn, Trickle:2019nya, Cox:2019cod, Griffin:2019mvc, Mitridate:2020kly, Trickle:2020oki, Griffin:2020lgd, Coskuner:2021qxo, Knapen:2021bwg, Mitridate:2023izi}. Because phonon dispersions and polarizations can vary with crystal direction, the DM-phonon excitation rate exhibits a daily modulation. This effect is primarily present for DM masses below 1 MeV and energy depositions below $\sim 100$ meV, which is currently below experimental thresholds.

In between the single-phonon and nuclear recoil limits, DM scattering is expected to produce multiphonon signals~\cite{Campbell-Deem:2022fqm, Campbell-Deem:2019hdx, Knapen:2020aky, Kahn:2020fef, Berghaus:2022pbu, Lin:2023slv, Schober2014, Kahn:2021ttr}. Here DM-nucleus scattering no longer behaves simply as a nuclear recoil, but instead the material response is broadened and additional features arising from the phonon density of states appear~\cite{Campbell-Deem:2022fqm}.
Experimental efforts are now underway to detect low-energy phonon-based signals~\cite{SPICE7,CRESST:2019jnq,SuperCDMS:2020aus}, and in the near future may achieve thresholds as low as $0.1-1$ eV, which sit in this multiphonon regime. It is therefore of interest to understand whether modulation signals are possible in multiphonons.

In this paper, we calculate modulation effects in multiphonons and show that they can provide another avenue for directional detection of nuclear recoils. This is a particularly appealing possibility for phonon-only signals, where backgrounds may be difficult to predict. In the presence of an unknown background, total rate measurements only provide upper limits and the sensitivity scales with the background rate. In contrast, modulation measurements can allow for background rejection and the potential for discovery.

We focus on a sapphire (Al$_2$O$_3$) target crystal, which is planned to be used in upcoming experiments~\cite{SPICE7} and known to give large anisotropies in DM-phonon excitations~\cite{Griffin:2018bjn}. We generalize the multiphonon calculation of Ref.~\cite{Campbell-Deem:2022fqm} to account for the anisotropic phonon density of states. We describe the calculation of the anisotropic crystal response in Sec.~\ref{sec:structurefactor}, reviewing some of the necessary approximations to obtain a tractable result. One key approximation, the incoherent approximation, limits our results to DM masses above $\sim $MeV. In Sec.~\ref{sec:modulation}, we give the DM scattering rate, as well as discuss the connection between the anisotropic structure factor and the DM rate modulation. We then give results in Sec.~\ref{sec:results} for different DM form factors and mediators. Some convergence tests for the results are given in App.~\ref{convergence_test_appendix}. The \textrm{DarkELF} implementation is summarized in App.~\ref{darkelf_appendix}. Additional results for SiC are provided in App.~\ref{additional_results_appendix}, and we also confirm that isotropic crystals such as Si and GaAs give negligible modulation.

\section{Structure factor \label{sec:structurefactor}}

DM scattering with a crystal lattice depends on the dynamic structure factor $S(\bfq,\omega)$, which describes the response of the target to momentum transfer $\bfq$ and energy transfer $\omega$. For a review that discusses the dynamic structure factor in the context of DM direct detection, see Ref.~\cite{Kahn:2021ttr}. For a crystal of volume $V$ and containing $N$ unit cells, the dynamic structure factor can be written as:
\begin{align}
    S(\bfq,\omega) = \frac{2 \pi}{V} \sum_f \bigg\vert \sum_{\bfl}^N \sum_{d=1}^{\mathfrak{n}} \bra{\Phi_f} f_{\bfl d} e^{i \bfq \cdot \mathbf{r}_{\bfl d}} \ket{0}\bigg\vert^2 \delta(E_f - \omega),
    \label{eq:structure_factor_defn}
\end{align}
where $f_{\bfl d}$ is a coupling strength with the atom at position $\mathbf{r}_{\bfl d}$, $\bfl$ is the lattice vector of a unit cell and $d$ labels the atom in the unit cell. This expression contains a sum over final states with energy $E_{f}$, so that each term in the sum gives the probability of the system to be excited to the state $\ket{\Phi_f}$. We work in the zero temperature limit, so that the initial state has no phonons.  

The dynamic structure factor has been evaluated for single-phonon excitations using first-principles phonon calculations~\cite{Knapen:2017ekk, Griffin:2018bjn, Trickle:2019nya, Cox:2019cod, Griffin:2019mvc, Mitridate:2020kly, Trickle:2020oki, Griffin:2020lgd, Coskuner:2021qxo, Knapen:2021bwg, Mitridate:2023izi}, while at high $\omega, q$ the dynamic structure is expected to reproduce elastic nuclear recoils. In between these regimes, multiphonon excitations are expected to dominate, but existing first principles techniques become very challenging when many phonons are produced. 
At low momentum, two-phonon excitations have been evaluated using the long-wavelength approximation and an effective theory of elastic waves~\cite{Campbell-Deem:2019hdx}. At higher momentum, a relatively simple result for the multiphonon contribution can be obtained with a few different approximations~\cite{Campbell-Deem:2022fqm,Schober2014}. These are the incoherent approximation, the assumption of a harmonic crystal, and the isotropic approximation. We next describe the role of each of these.

The first assumption is the incoherent approximation, which applies when the momentum transfer is greater than the inverse lattice spacing, $q \gtrsim 2 \pi / a$. Here we neglect interference terms between atoms and assume that the scattering off individual atoms dominates \eqref{eq:structure_factor_defn}. Mathematically, we compute the sum of the squared amplitudes rather than the square of the summed amplitudes over all atoms. We approximate the total structure factor as~\cite{Campbell-Deem:2022fqm}:
\begin{align}
        \label{eq:inc-approx}
        S(\bfq,\omega) \approx \sum_{\bfl}^N \sum_d^{\mathfrak{n}} \overline{f_d^2}\mathcal{C}_{\bfl d},
\end{align}
where the auto-correlation function is given by
\begin{equation}
    \label{eqn:incoherent_Cld}
    \mathcal{C}_{\bfl d} = \frac{1}{V} \int\displaylimits_{-\infty}^{\infty} \! dt\, e^{-2 W_d(\mathbf{q})} e^{\langle \mathbf{q} \cdot \mathbf{u}_{\bfl d}(0) \, \mathbf{q} \cdot \mathbf{u}_{\bfl d} (t) \rangle} e^{-i\omega t},
\end{equation}
and the Debye-Waller factor is
\begin{align}
     W_d (\bfq) &= \frac{1}{2} \langle ( \bfq \cdot \bfu_{\bfl d}(0) )^{2} \rangle.
\end{align}
Here, $\bfu_{\bfl d}$ is the displacement of an atom from equilibrium such that $\mathbf{r_{\bfl d}} = \bfl + \mathbf{r}_d^0 + \bfu_{\bfl d}$. The coupling strength has been replaced by the quantity $\overline{f_d^2}$, which is the average of the coupling quantity squared over different atoms of type $d$ in the crystal. We will only consider pure crystals with a single isotope for each atom, so we can assume that $(\overline{f_{d}})^2 = \overline{f_d^2}$ for all calculations, however we have left results in terms of $\overline{f_d^2}$. For mediators coupling equally to protons and neutrons, we will take $f_d^2 = A_d^2$.

The second approximation is to assume a harmonic crystal, which means that the phonon Hamiltonian is quadratic and that higher-order anharmonic phonon interactions are neglected. The utility of this approximation is that it allows us to write $\mathcal{C}_{\bfl d}$ in terms of the phonon density of states. The validity of the harmonic approximation was subsequently studied in Ref.~\cite{Lin:2023slv} and it was shown to be an excellent approximation over most of the scattering phase space. We continue to make this assumption.

Finally, Ref.~\cite{Campbell-Deem:2022fqm} also assumed an isotropic crystal. In general, the phonon energies and eigenvectors vary along different directions in the crystal. Neglecting this variation, the result can be simplified to depend only on an averaged density of states. This is expected to be an excellent approximation for materials such as Si and GaAs, and it was not known how well it would hold for more anisotropic crystals.

In the remainder of the section, we review the calculations of the structure factor in the isotropic approximation, and then discuss the generalization for anisotropic crystals in terms of a phonon density of states tensor.

\subsection{Isotropic approximation}

Following \cite{Campbell-Deem:2022fqm}, in the harmonic approximation, we can write the displacement vector in a phonon mode expansion
\begin{align}
    \label{eq:u_ld(t)}
    \bfu_{\bfl d} (t) = \sum_{\nu} \sum_{\bfk} & \frac{1}{\sqrt{2 N m_d \omega_{\nu , \bfk}}}(\bfe_{\nu , d, \bfk} \hat{a}_{\nu, \bfk}e^{i \bfk \cdot (\bfl + \mathbf{r}_d^0) - i \omega_{\nu , \bfk} t} + \nonumber \\
    &\bfe^*_{\nu , d, \bfk} \hat{a}^{\dagger}_{\nu, \bfk}e^{-i \bfk \cdot (\bfl + \mathbf{r}_d^0) + i \omega_{\nu , \bfk} t}), 
\end{align}
where $\nu$ denotes the phonon branches, $\bfk$ is the phonon momentum in the first Brillouin Zone (BZ), $\hat{a}^{\dagger}_{\nu, \bfk}$ and $\hat{a}_{\nu, \bfk}$ are the creation and annihilation operators, and $\bfe_{\nu, d, \bfk}$ are the phonon eigenvectors with associated energies $\omega_{\nu, \bfk}$.

To evaluate the structure factor in the incoherent approximation,~\eqref{eq:inc-approx}, we first evaluate the correlator $\langle \bfq \cdot \bfu_{d}(0) \, \bfq \cdot \bfu_{d}(t) \rangle$ using \eqref{eq:u_ld(t)}:
\begin{align}
    \langle \bfq \cdot \bfu_{d}(0) \, \bfq \cdot \bfu_{d}(t) \rangle = q^{2} \sum_{\nu} \sum_{\bfk} \frac{ |\hat \bfq \cdot \bfe_{\nu, \bfk, d} |^2 }{2 N  m_d \omega_{\nu, \bfk}}  e^{i \omega_{\nu,\bfk} t}.
\end{align}
The quantity on the right hand side is related to the phonon density of states. To see this, we work in the isotropic approximation and average over the direction of $\hat{\bfq}$, yielding
\begin{align}
    &\langle \bfq \cdot \bfu_{d}(0) \, \bfq \cdot \bfu_{d}(t) \rangle \approx 
    \frac{q^2}{3} \sum_{\nu} \sum_{\bfk} \frac{ |\bfe_{\nu, \bfk, d} |^2 }{2 N  m_d \omega_{\nu, \bfk}}  e^{i \omega_{\nu,\bfk} t} \\
    &= \frac{q^2}{3}  \sum_{\nu} \sum_{\bfk} \int_{-\infty}^{\infty} d \omega' \delta (\omega ' - \omega_{\nu, \bfk}) \frac{ |\bfe_{\nu, \bfk, d} |^2 }{2 N  m_d \omega'}  e^{i \omega' t}\\
    &= \frac{q^2}{2 m_d} \int_{-\infty}^{\infty} d \omega' \frac{D_d(\omega')}{\omega '} e^{i \omega ' t}
\end{align}
where we have defined the partial density of states,
\begin{align}
    D_d(\omega) = \frac{1}{3N} \sum_{\nu} \sum_{\bfk} |\bfe_{\nu, \bfk, d} |^2 \delta (\omega - \omega_{\nu, \bfk}),
\end{align}
which is normalized to satisfy $\int_{-\infty}^{\infty} d \omega D_d(\omega) = 1$. Similarly, here, the isotropic Debye-Waller factor takes the form:
\begin{align}
    W_d (\bfq) \approx \frac{q^2}{4 m_d} \int_{-\infty}^{\infty} d \omega' \frac{D_d(\omega ')}{\omega ' }. 
\end{align}

Now, we can evaluate the auto-correlation function \eqref{eqn:incoherent_Cld} in a phonon number expansion. That is, we take a series expansion of $e^{\langle \bfq \cdot \bfu_{d}(0) \, \bfq \cdot \bfu_{d}(t) \rangle}$ in powers of $q^{2n}$ for $n$ phonons being excited. From this we have the full isotropic structure factor (in the incoherent, harmonic approximation):
\begin{align}
    \label{eq:isotropic_Sqw}
    S(\bfq,\omega) & \approx \frac{2 \pi}{\Omega} \sum_d^{\mathfrak{n}} \overline{f_d^2} e^{-2 W_d (q)} \sum_n \frac{1}{n!} \left(\frac{q^2}{2 m_d} \right)^n \\
    & \times \left(\prod_{i=1}^n \int d\omega_i \frac{D_d(\omega_i)}{\omega_i} \right) \delta \left(\sum_j \omega_j - \omega \right) \nonumber.
\end{align}
Here we have defined $\Omega = V/N$ as the volume per unit cell. The $n$th term then consists of an integral over individual phonon energies weighted by the density of states, and with the total energy equal to $\omega$.

As $q$ increases, the higher-$n$ terms contribute more in the structure factor. Since $n! \propto \sqrt{n}(\frac{n}{e})^n$ at large $n$, the $n$th term will start to contribute when $\frac{q^2}{2 m_d \overline{\omega_d}} \sim n$, where $\overline{\omega_d} = \int d \omega' \omega' D_d (\omega')$ is a typical phonon energy. When $q$ becomes sufficiently large that the typical number of phonons $n \gg 1$, the total structure factor approaches a Gaussian. To speed up calculations and avoid evaluating sums over many phonons, it is also useful to employ the impulse approximation (IA). This can be obtained by approximating \eqref{eqn:incoherent_Cld} with the saddle point at $t=0$, corresponding to an impulse, which gives
\begin{align}
    \label{eq:impulse_approx}
    S^{IA}(\mathbf{q},\omega) = \sum_d \frac{\overline{f_d^2}}{\Omega}  \sqrt{\frac{2 \pi}{\Delta_d^2}}\exp{ \left(-\frac{(\omega - \frac{q^2}{2m_d})^2}{2 \Delta_d^2} \right)}
\end{align}
with width
\begin{align}
    \Delta_d^2 = \frac{q^2 \overline{\omega_d}}{2 m_d}.
\end{align}
We use this result for $q > 2 \sqrt{2 m_d \omega_0}$ with $\omega_0 = \frac{1}{\overline{\omega_d^{-1}}}$ where
\begin{align}
    \overline{\omega_d^{-1}} = \int d \omega' \frac{D_d (\omega ')}{\omega '}.
\end{align}
While $\omega_0$ is $d$-dependent, for multi-atomic lattices, we use $d$ associated with the most massive atom. For other quantities such as the width $\Delta_d^2$, and whenever $d$ is explicitly noted, we include all atoms and account for the atom dependence.

As discussed in Ref.~\cite{Campbell-Deem:2022fqm}, this result allows for an explicit connection between multiphonon processes and the nuclear recoil limit. For $q \gg 2 \sqrt{2 m_d \omega_0}$, approximating the Gaussian as a delta-function leads to a scattering rate which matches the usual result for nuclear recoils.

\subsection{Anisotropic correlation function}

 \begin{figure*}
     \centering
     \includegraphics[width=1\linewidth]{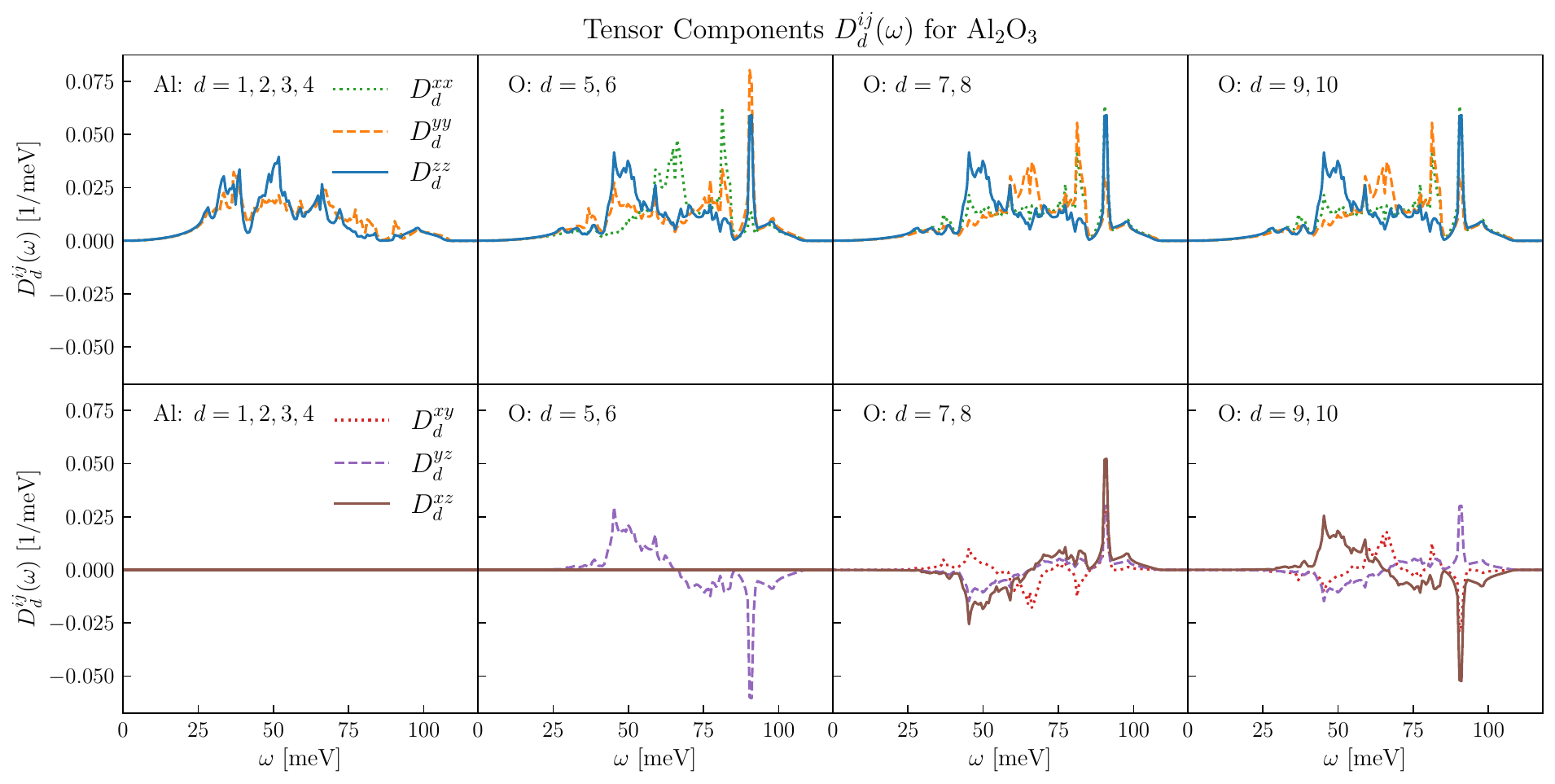}
     \caption{Density of states tensor for Al$_2$O$_3$, where redundant atoms with identical density of states are indicated. The diagonal and off-diagonal components are shown in the top and bottom row, respectively. For spin-independent scattering, Al dominates, making the anisotropies in O less important.} 

     \label{fig:pdos_Al2O3}
 \end{figure*}

To evaluate the anisotropic structure factor, we return to the correlator $\langle \bfq \cdot \bfu_{d}(0) \, \bfq \cdot \bfu_{d}(t) \rangle$.
Rather than averaging over the direction of $\bfq$ as in the isotropic case, we can consider projections onto specific directions of $\bfq$
\begin{align}
\langle \bfq \cdot \bfu_{d}(0) \, \bfq \cdot \bfu_{d}(t) \rangle &= q^{2} \sum_{\nu} \sum_{\bfk} \frac{ |\hat \bfq \cdot \bfe_{\nu, \bfk, d} |^2 }{2 N  m_d \omega_{\nu, \bfk}}  e^{i \omega_{\nu,\bfk} t}\\
&= \frac{ q^2}{2 m_d}\int\displaylimits_{-\infty}^{+\infty}\!\! d\omega'\frac{D^{\hat q}_d(\omega') }{\omega'}  e^{i \omega' t}.
\label{eq:correldensitystates}
 \end{align}
Now the projected density of states (pDoS) is defined as
\begin{align}
	D^{\hat q}_d(\omega) = \frac{1}{N}  \sum_{\nu}  \sum_{\bfk} |\hat \bfq \cdot \bfe_{\nu, \bfk, d} |^2 \delta(\omega - \omega_{\nu, \bfk} ).
\end{align}
The normalization is again $\int_{-\infty}^{\infty} d \omega D^{\hat q}_d(\omega) = 1$.
The Debye-Waller factor also needs to be modified, where now
\begin{align}
	W_{d}(\bfq) &= \frac{ q^2}{4 m_d}\int_{-\infty}^{+\infty}\!\! d\omega'\frac{D^{\hat q}_d(\omega') }{\omega'}.
 \end{align}
Given the pDoS, we can evaluate the full DM scattering rate without the isotropic approximation.

The phonon pDoS can be calculated for a given direction $\hat q$ using the public code \texttt{phonopy}~\cite{togo2015principles}, combined with the relevant calculations of second-order force constants. DM scattering rates involve an integral over all directions of $\hat q$, and it is not practical or necessary to redo the pDoS calculation for every $\hat q$. Instead, we can capture all the relevant information inside a density of states tensor defined as
\begin{align}
    \label{eq:dos_tensor}
	\mathbf{D}^{ij}_d(\omega) = \frac{1}{N}  \sum_{\nu}  \sum_{\bfk}  (\hat r_{i} \cdot \bfe_{\nu, \bfk, d} )( \bfe_{\nu, \bfk, d}^{*} \cdot \hat r_{j} ) \, \delta(\omega - \omega_{\nu, \bfk} ).
\end{align}
where $\hat r_{i} = \hat x, \hat y, \hat z$. Decomposing this 3$\times$3 tensor into its symmetric and anti-symmetric parts, $\mathbf{D}^{ij} = D^{ij} + A^{ij}$, we can decompose any arbitrary $D^{\hat q}_d(\omega)$ as
\begin{align}
	D^{\hat q}_d(\omega)  &= \sum_{i,j} \hat q_{i} \mathbf{D}^{ij}_d(\omega) \hat q_{j} \\
    &= \sum_{i,j} \hat q_{i} D^{ij}_d(\omega) \hat q_{j}.
\end{align}
While $\mathbf{D}^{ij}_d(\omega)$ may not be symmetric due to off-diagonal complex contributions in \eqref{eq:dos_tensor}, the quantity $\hat q_{i} \mathbf{D}^{ij}_d(\omega) \hat q_{j}$ only depends on its symmetric part, so we only need to find $D^{ij}_d(\omega)$, which is a real and positive definite tensor.

\begin{figure*}
    \centering
    \subfloat[ ][Mode 9, $\omega = 50$ meV, low $\bfq$ along $\hat z$.]{
        \includegraphics[width=0.21\linewidth]{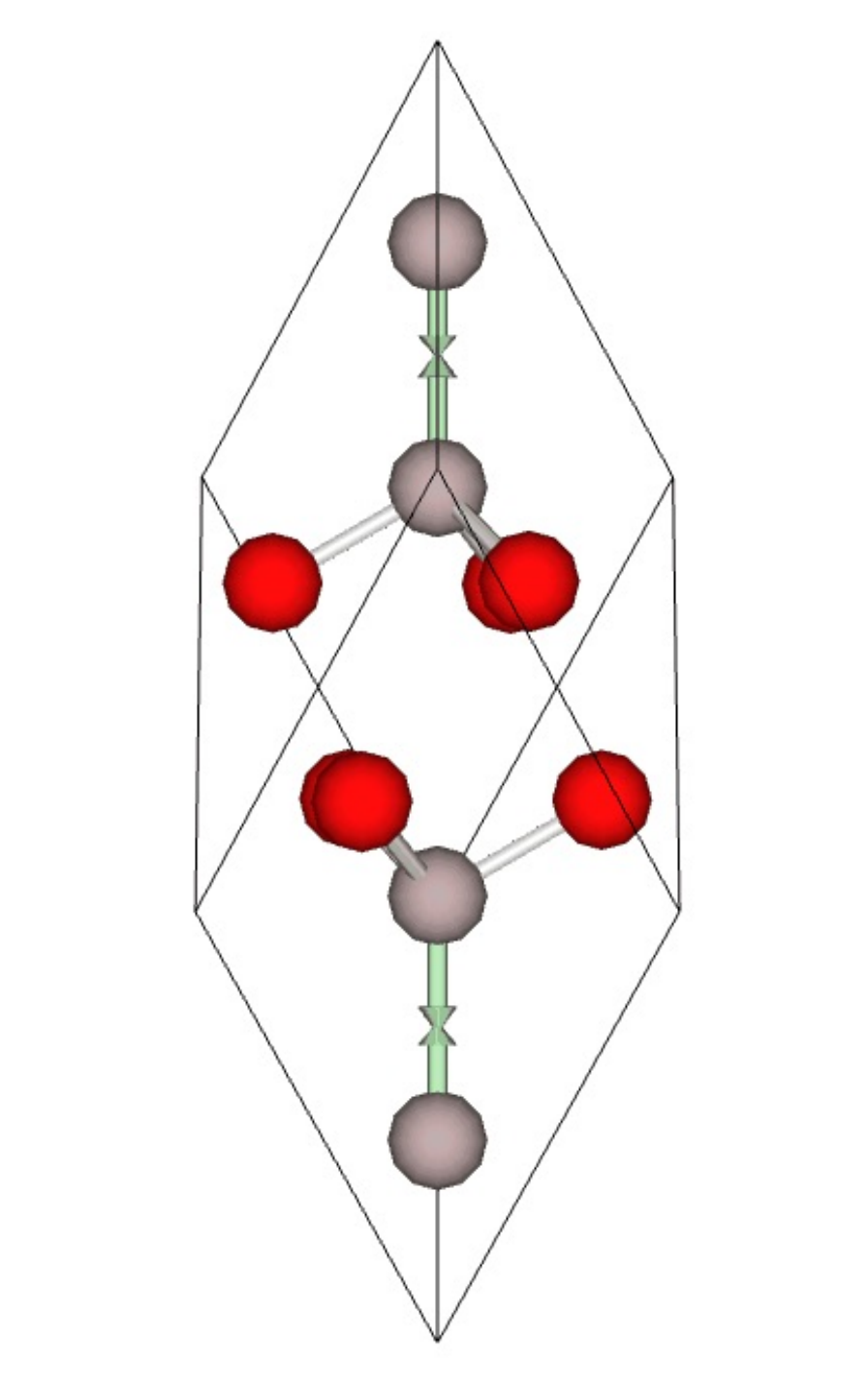}
        }
    \hspace{0.5cm}
    \subfloat[ ][Mode 17, $\omega = 50$ meV, medium $\bfq$ along $\hat{z}$.]{
        \includegraphics[width=0.19\linewidth]{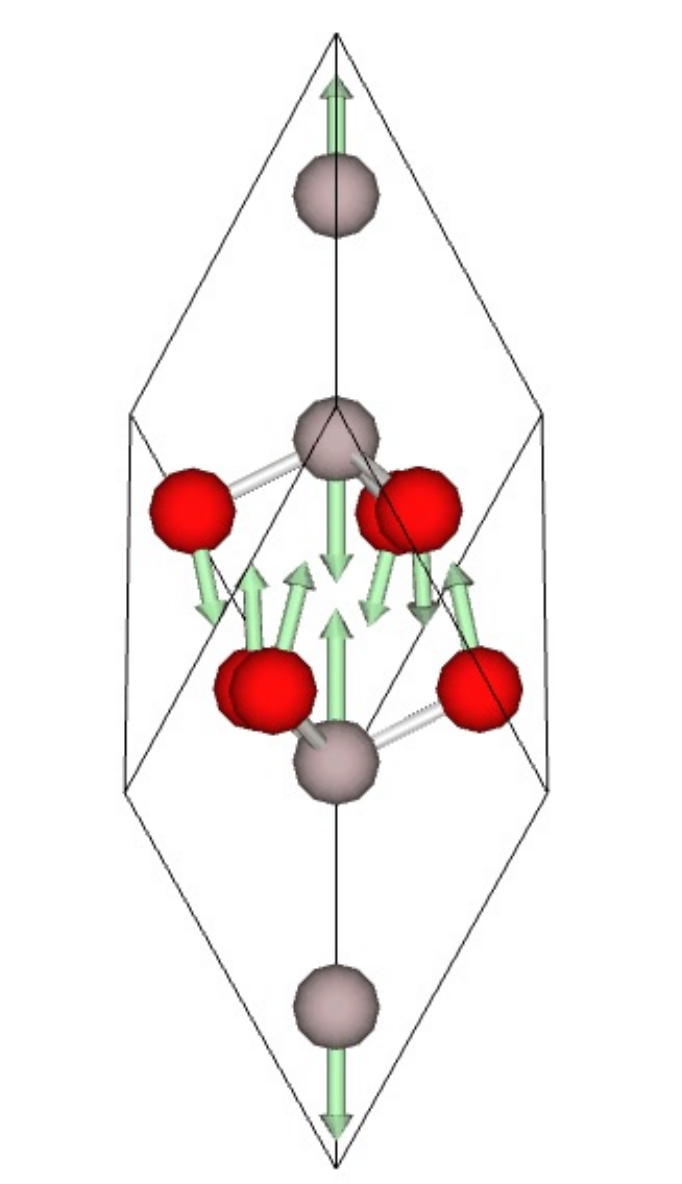}
        } 
    \hspace{0.5cm}
    \subfloat[ ][Mode 26, $\omega = 78$ meV, low $\bfq$ along $\hat{y}$.]{
        \includegraphics[width=0.205\linewidth]{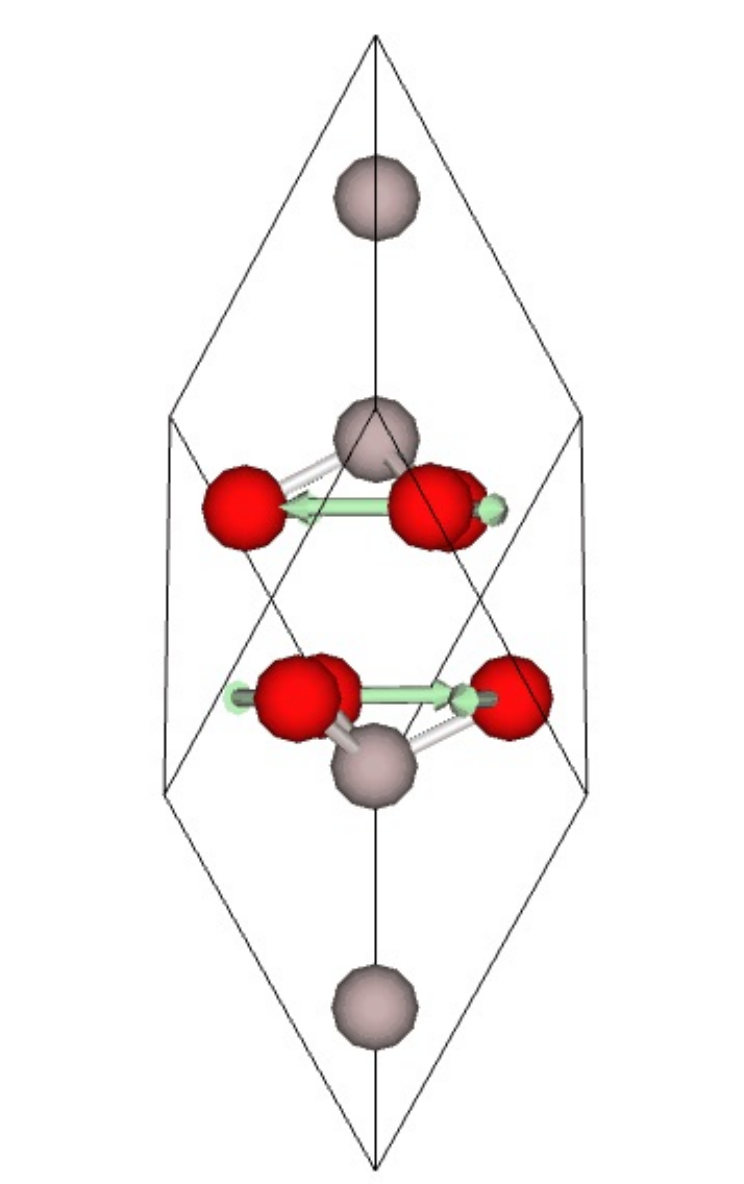}
        }
    \caption{Selected phonon eigenmodes for sapphire corresponding to directional peaks in pDoS, viewing the crystal from the $\hat{y}$ axis, with the $\hat{z}$ axis shown vertically. Al atoms are gray, and O atoms are red. Images are generated with \cite{phononwebsite}.}
    \label{fig:al2o3_visual_modes}
\end{figure*}

We reconstruct the tensor $D^{ij}_d(\omega)$ from \texttt{phonopy} by evaluating $D^{\hat q}_d(\omega)$ for different directions. We evaluate the diagonal components directly by calculating pDoS along $\hat q = \hat x, \hat y, \hat z$. To obtain the off-diagonal components, we evaluate $D^{\hat q}_d(\omega)$ for $\hat q = \frac{1}{\sqrt{2}} ( \hat x + \hat y)$, which can be written as
\begin{align}
    \label{eq:off_axis_dos_tensor}
	D^{\hat q}_d(\omega) = \frac{1}{2} \left( D^{xx}_{d}(\omega) + D^{yy}_{d}(\omega) + 2 D^{xy}_{d}(\omega) \right)
\end{align}
allowing us to extract $D^{xy}_{d}(\omega)$. Repeating this procedure, we can extract all off-diagonal components.

In Figure \ref{fig:pdos_Al2O3}, we show the components of the pDoS tensor for Al$_2$O$_3$, which has a rhombohedral crystal structure with a primary axis of symmetry. The force constants used are those provided with \texttt{phonopy}. We use the standard coordinate system with $\hat{z}$ along the primary axis. Al$_2$O$_3$ has 4 Al atoms (labeled $d=1,...4$) and 6 O atoms (labeled $d = 6,...10$) in the unit cell. As expected, due to the anisotropic crystal, there is significant variation across different directions $\hat{q}$ and non-zero off-axis components $D_d^{ij}$, $i \neq j$. There is also a redundancy in the tensor components between different atoms $d$. All of the Al atoms are equivalent, and three distinct pairs of O atoms are equivalent.

Sapphire has 30 phonon eigenmodes, which can display different motion among the 10 atoms in the unit cell at different energies $\omega$ and momentum transfer $\bfq$, making it difficult to analyze the directional pDoS in terms of specific modes. However, we can connect a few modes to domminant features of the pDoS. In Figure~\ref{fig:al2o3_visual_modes}, using \cite{phononwebsite} we select three phonon modes to visualize that exhibit relatively coherent motion to explain some of the peaks in the components of the $D_d^{ij}$ tensor. In the first two columns, we show mode 9 and mode 17 at $\omega = 50$ meV, respectively, with $\bfq$ along $\hat z$. In the pDoS, there are peaks in all atoms along $D_d^{zz}$ at $\omega \sim 50$ meV. This can be partially attributed to these longitudinal optical (LO) modes with motion mainly restricted along $\hat z$. Other modes at $\omega \sim 50$ meV with $\bfq$ along the crystal's primary axis demonstrate similar motion along $\hat z$, while $\bfq$ along the $xy$ plane at this energy results in incoherent, random motion. On the right, we show mode 26 at a low $\bfq$ along $\hat y$ at $\omega = 78$ meV. Here, this mixed TO-LO mode (TO transverse optical) with motion in just the O atoms helps explain the peaks in $D_d^{xx}$ and $D_d^{yy}$ around this $\omega$ that exist only for the O atoms. For different $\bfq$ at this energy, O atoms also show motion along $xy$ and there is very little motion among the Al atoms, explaining the overall decrease in the pDoS for Al.

Finally, the full result for the anisotropic structure factor is given by
\begin{align}
    \label{eq:anisotropic_Sqw}
	& S(\bfq, \omega) \approx \frac{2 \pi}{\Omega}   \sum_{d}^{\mathfrak{n}}  \overline{f_d^2} \,  e^{-2 W_{d}({\bf q})} \times \\
	& \sum_{n} \frac{1}{n!} \left( \frac{q^{2}}{2 m_{d}} \right)^{n}  \left(\prod\displaylimits_{i=1}^n \int d \omega_{i} \frac{D^{\hat q}_d(\omega_i)}{\omega_i} \right) \delta \left(\sum_j \omega_j - \omega \right), \nonumber
\end{align}
where we have replaced the isotropic Debye-Waller factor and partial density of states in \eqref{eq:isotropic_Sqw} with the direction-dependent version.

Inside both isotropic and anisotropic structure factors, there is an $n$-dimensional integral for the $n$-phonon term, defined for the anisotropic case as 
\begin{align}
\label{eq:T_n(omega)}
T_{n}^{\hat q}(\omega) = \left(\prod\displaylimits_{i=1}^n \int d \omega_{i} \frac{D^{\hat q}_d(\omega_i)}{\omega_i} \right) \delta \left(\sum_j \omega_j - \omega \right).
\end{align}
It is helpful to avoid directly evaluating this integral, and instead use the recursion relation 
\begin{align}
T_{n}^{\hat q}(\omega) = \int d\omega_{n} T_{1}^{\hat q}(\omega_{n}) T_{n-1}^{\hat q}(\omega - \omega_{n}) 
\end{align}
where 
\begin{align}
T_{1}^{\hat q}(\omega) = \frac{D_{d}^{\hat q}(\omega)}{\omega}.
\end{align}
This speeds up calculations substantially.

As in the isotropic case, the anisotropic structure factor approaches a Gaussian at high $q$. We again make use of the impulse approximation in Eq.~\eqref{eq:impulse_approx}, now using a higher threshold of $q > 4 \sqrt{2 m_N \omega_0}$. We could evaluate the impulse approximation using the direction-dependent $D_d^{\hat q}(\omega)$. However, since the Gaussian is expected to approach the nuclear recoil limit, the effects of anisotropy are expected to subside for higher $q$. The direction-dependence only appears in the width of the Gaussian $\Delta_d^2$, whereas in the nuclear recoil limit the total rate can be well-approximated by taking the Gaussian to a delta-function. Therefore we will use the isotropic approximation for impulse approximation calculations, choosing $\hat{q}=\hat{z}$ in calculating the width of the Gaussian $\Delta_d^2$. For Al$_2$O$_3$, $\overline{\omega}_d$ is only 6-8\% higher along $xy$ than along $\hat z$, depending on the atom, leading to a mild change in $\Delta_d^2$.

\begin{figure*}
    \centering
    \includegraphics[width=1\linewidth]{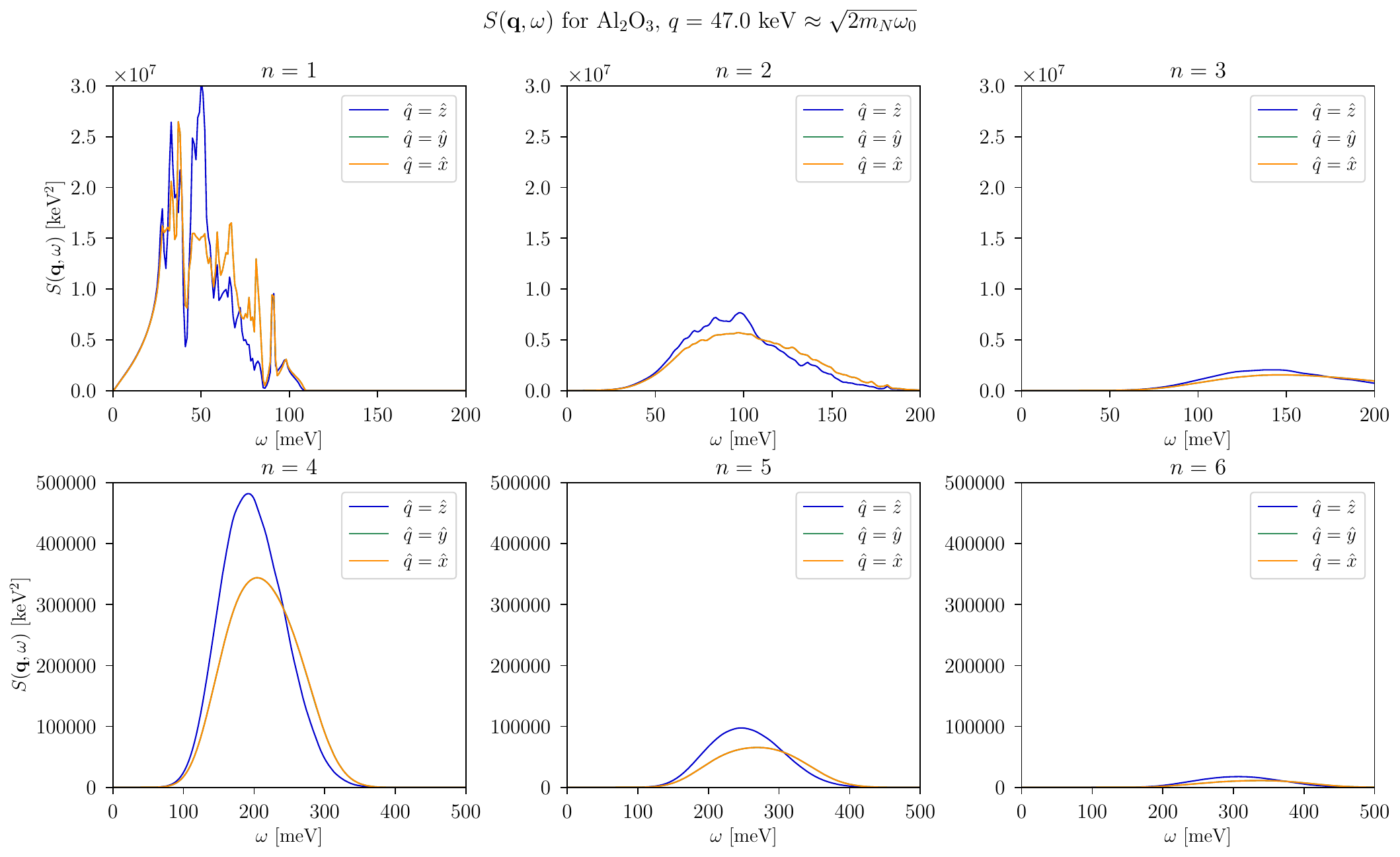}
    \caption{Decomposition of the anisotropic structure factor into individual phonon terms, for Al$_2$O$_3$ and at fixed $q = 47$ keV. The structure factor is computed assuming $f_d = A_d$.  We show results for $\hat{q}= \hat x, \hat y, \hat z$. The lines for $\hat x$ and $\hat y$ are identical, implying rotational symmetry with respect to $\hat z$. For all $n$, there is a stronger response for $\hat q = \hat z$ at lower $\omega$, while at higher $\omega$ it is stronger for the $xy$ plane. 
    Note that the axis scales for the top three panels are different from the bottom three.}
    \label{fig:Al2O3_Sqw_nphonon}
\end{figure*}

\begin{figure*}
    \centering
    \includegraphics[width=1.0\linewidth]{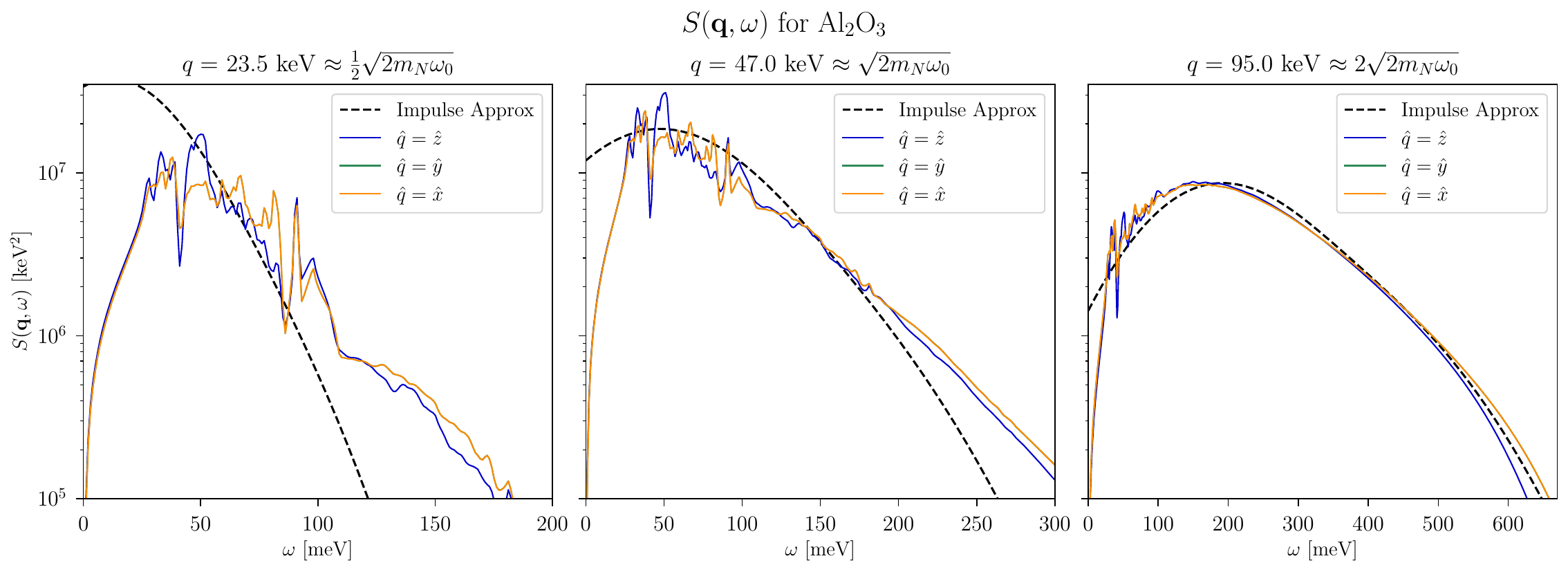}
    \caption{The anisotropic structure factor for Al$_2$O$_3$ for three fixed $q$ values, $\hat{q}$ evaluated at $\hat x, \hat y, \hat z$, and summing over $n=1...20$. Like Fig.~\ref{fig:Al2O3_Sqw_nphonon}, the lines associated with $\hat x$ and $\hat y$ are identical. For all $q$, there are regions in $\omega$ where certain directions $\hat q$ dominate; these regions are most pronounced for lower $q$, as in the left two plots. The most notable region is $\omega \sim 50$ meV, where $\hat z$ dominates. For $ 60 \lesssim \omega \lesssim 85$ meV, the $xy$ plane dominates, but to a lesser extent. In general, the high $\omega$ tail has a suppression in the $\hat q = \hat z$ direction due to the wider tails in the individual phonon terms along $xy$. At high $q$ (right panel), the directional dependence tends to smooth out as the response approaches the nuclear recoil limit.
    Also shown is the impulse approximation, which becomes a good approximation for $q \gtrsim 2 \sqrt{2 m_N \omega_0}$.}
    \label{fig:Al2O3_Sqw}
\end{figure*}

\subsection{Numerical Results}

In the isotropic approximation, the structure factor gives material specific information, dictating what energy transfers $\omega$ and momentum transfers $q$ contribute the largest material response. This is useful for understanding how the energy threshold $E_{thresh}$ of a detector affects the observed rate. In the anisotropic case, $S(\bfq,\omega)$ also offers insight into the directional dependence on $\bfq$, which can create a modulating daily rate as the crystal rotates relative to the DM wind.

To understand how the structure factor varies for different directions $\hat{q}$ and at different $\omega$, we first decompose $S(\bfq,\omega)$ into individual phonon terms in Fig.~\ref{fig:Al2O3_Sqw_nphonon}. We fix the magnitude of the momentum transfer to be $q \approx \sqrt{2 m_d \omega_0}$, where $m_d$ and $\omega_0$ are for Al and $\omega_0$ along $\hat q = \hat z$. We also take the coupling strength $f_d = A_d$ for equal couplings to nucleons. Note that in this case that scattering off Al dominates due to the larger $A_d$, and the structure factor is $\sim 3$ times higher for Al at small $q$.

First consider the $n=1$ panel of Fig.~\ref{fig:Al2O3_Sqw_nphonon}. This directional dependence is a result of the lattice structure of the target crystal, and arises directly from the pDoS shown in Fig.~\ref{fig:pdos_Al2O3}. The most noticeable direction dependence at $\omega \sim 50$ meV, favoring the $\hat z$ direction, is explained by the same anisotropy in the pDoS for both Al and O at that energy. For $60 \lesssim \omega \lesssim 85$ meV, the pDoS is instead highly peaked in the $\hat x$ and $\hat y$ directions for the O atoms. These individual features become much harder to see for $n \ge 2$, but these differences in the pDoS show up in the stronger response for $\hat z$ at lower $\omega$ and stronger response along $xy$ at higher $\omega$.
There is also a rotational symmetry with respect to the $\hat z$ axis. 
Due to the rhombohedral structure of Al$_2$O$_3$, it has a symmetry with respect to its $z$-axis, which suggests an explanation for the difference between scattering along $\hat{q} = \hat{z}$ and the $x$-$y$ axis.

In Fig.~\ref{fig:Al2O3_Sqw}, we show the total structure factor when summing from $n=1,...,10$.
Successive plots show the evolution for several values of $q$, and all plots consider three directions, $\hat{q} = \hat x, \hat y, \hat z$.  The clearest effects of anisotropy are seen in the left two plots for lower $q \lesssim 2 \sqrt{2 m_N \omega_0}$. The most dramatic region where $S(\bfq,\omega)$ shows directional dependence is $\omega \sim 50$ meV, where the response is stronger for $\hat q = \hat z$.  This feature is inherited from the $n=1$ term in the sum. Similarly, the next most dramatic region of $60 \lesssim \omega \lesssim 85$ meV, where the $xy$ plane dominates, is also explained by the differences in the $n=1$ term.

As $q$ increases in subsequent plots, multiphonon contributions become more substantial and anisotropies are less noticeable.
Although there are clear differences for $\hat{q} = \hat z$ versus the $xy$ plane for individual $n$  in Fig.~\ref{fig:Al2O3_Sqw_nphonon}, these become much less prominent when summing over the $n$-phonon terms. For a given $n$, there is a relatively smaller response for $\hat{q} = \hat z$ at higher $\omega$; however, this is partially compensated for when we consider that the $n + 1$ term has a relatively larger response in the same range of $\omega$. This washing out of the anisotropy at high $q$ is consistent with the expectation that the structure factor approaches the impulse approximation, indicated by the dashed line in Fig.~\ref{fig:Al2O3_Sqw}.

We note one more source of anisotropy. At very high $\omega$ in all panels, there is a slightly stronger response for the $xy$ plane. This is normally inconsequential, except when instituting a relatively high threshold on the $\omega$ integration. This slight suppression of the $\hat z$ direction is due to the tails of individual $n$ terms weighted towards $xy$.

\section{Origin of daily modulation \label{sec:modulation}}

Having understood the features in the anisotropic structure factor, we now explore how it leads to daily modulation in the DM scattering rate. 

We first give the full DM scattering rate in terms of the structure factor:
\begin{align}
\label{eq:total_rate}
R(t) = \frac{1}{\rho_{T}} \frac{\rho_{\chi}}{m_{\chi}} \int \frac{d^{3} \bfq}{(2\pi)^{3}} d \omega \, g(\bfq, \omega, t) \frac{\pi \sigma(q)}{\mu_{\chi}^{2}} S(\bfq, \omega)
\end{align}
and the integration over the DM velocity distribution can be written as
\begin{align}
 g(\bfq, \omega, t) = \int d^{3} \bfv f( \bfv + \bfv_{e}(t)) \delta(\omega - \omega_{\bfq})
\end{align}
where $\omega_{\bfq} = \bfq \cdot \bfv - q^{2}/(2m_{\chi})$ and the Earth's velocity is given by
\begin{align}
	\mathbf{v}_e (t) = v_e 
    \begin{pmatrix} 
        \sin{\theta_e} \sin{\phi_e(t)} \\
        \sin{\theta_e} \cos{\theta_e} (\cos{\phi_e(t) - 1}) \\
        \cos^2{\theta_e} + \sin^2{\theta_e}\cos{\phi_e(t)} 
    \end{pmatrix}
\end{align}
as in \cite{Griffin:2018bjn, Coskuner:2019odd}. 
We assume $v_e = 240$ km/s, $\theta_e = 42^{\circ}$ and $\phi_e(t) = 2 \pi (\frac{t}{24 \, \text{hours}})$.  $\rho_T$ is the target density, $\rho_{\chi} = 0.4$ GeV/cm$^3$ is the local DM density and $\mu_{\chi}$ is the DM-nucleon reduced mass. This setup assumes that at $t~=~0$, the $z$-axis of the crystal is aligned with the Earth's velocity $\bfv_e$, and is independent of the lab latitude. All variables are thus evaluated in the frame of the crystal in Eq.~$\eqref{eq:total_rate}$.

The momentum dependent cross section is $\sigma(q) = \sigma_0 |\Tilde{F}(q)|^2$, for a reference cross section $\sigma_0$ and form factor $\Tilde{F}(q) = \frac{q_0^2 + m_{med}^2}{q^2 + m_{med}^2}$ with model-dependent reference momentum $q_0$. While this form factor is dependent on the mediator mass $m_{med}$, we consider two limiting cases. In the massive mediator limit, $\Tilde{F}(q) = 1$, and in the massless mediator limit, $\Tilde{F}(q) = \frac{q_0^2}{q^2}$.

As the Earth rotates, the orientation of the detector with respect to the Earth's velocity $\mathbf{v}_e$ changes. Unlike isotropic crystals where the target response is independent of the direction of momentum transfer, the variation in the structure factor of an anisotropic crystal across different $\hat q$ can lead to a daily modulation in the overall rate.
In this section, we use a heavy scalar mediator with $f_d^2 = A_d^2$ and $|\Tilde{F}(q)|^2 = 1$ to illustrate sources of daily modulation. The physics is qualitatively similar for other mediators, which will be shown in Sec.~\ref{sec:results}.

Whereas $S(\bfq,\omega)$ gives target-specific information, the kinematic function $g(\bfq,\omega,t)$  dictates which $\omega$ and $\bfq$ contribute most to the rate as the Earth rotates. If the DM velocity distribution is given by the Standard Halo Model, the integral can be explicitly evaluated, see Eqs 11-12 of \cite{Coskuner:2021qxo},
\begin{align}
    \label{eq:g(q,w,t)_explicit}
    g(\bfq,\omega,t) = \frac{\pi v_0^2}{N_0 q} [\exp{(- \frac{v_{-} (\mathbf{q},\omega,t)^2}{v_0^2})} - \exp{(- \frac{v_{esc}^2}{v_0^2}})],
\end{align}
with $N_0$ a normalization constant and
\begin{align}
    v_{-} (\mathbf{q},\omega,t) = \min{(v_{esc},|\hat{q} \cdot \mathbf{v}_e (t) + \frac{q}{2 m_{\chi}} + \frac{\omega}{q} |)}.
\end{align}
We take $v_{esc} = 500$ km/s and $v_0 = 220$ km/s.

A few observations about the kinematic function. We can see that $g(\mathbf{q},\omega,t)$ is only non-zero when $|\hat{q} \cdot \mathbf{v}_e (t) + \frac{q}{2 m_{\chi}} + \frac{\omega}{q} | < v_{esc}$. Defining
\begin{align}
    v_{*} \equiv \frac{q}{2 m_{\chi}} + \frac{\omega}{q},
\end{align}
we can identify the regions where the kinematic function is non-zero as those satisfying
\begin{align}
    v_{*} < v_{esc} - \hat{q}\cdot \bfv_e(t).
\end{align}
If $v_{*}> v_{esc} + v_e$, then $g(\bfq,\omega,t)$ will always be zero at any time $t$ and direction $\hat{q}$. Similarly, if $v_{*} < v_{esc} - v_e$,  then $g(\bfq,\omega,t)$ will be nonzero for all $t$ and $\hat q$. That leaves the sensitive middle region
\begin{align}
    \label{eq:sensitive_v*}
    v_{*} &\in (v_{esc} - v_e, v_{esc} + v_e) 
    \approx (0.87, 2.47) \times 10^{-3}
\end{align}
where $g$ is sometimes zero, depending on the direction $\hat{q}$ and time $t$. 
For multiphonon rates, $q \gtrsim$ keV and $\omega \sim$ 100 meV, so $v_{*}$ is well approximated as $v_{*} \approx q / 2 m_\chi$, meaning that $g(\bfq,\omega,t)$ is effectively independent of $\omega$.

At the same time, the anisotropies in the structure factor are most prominent for $q \lesssim \sqrt{2m_d \omega_0}$. Therefore, setting $v_* \sim \sqrt{2 m_d \omega_0}/(2 m_\chi)$ and requiring $v_* > v_{esc} - v_e$, we can obtain an upper limit on $m_{\chi}$ where we expect to see noticeable daily modulation.
For sapphire, this upper mass limit is $m_{\chi} \approx 30$ MeV. This suggests that we will see the greatest impacts of anisotropy on multiphonon excitations in the mass range $m_\chi \approx 1- 30$ MeV.

\begin{figure*}
    \centering
    \includegraphics[width=1\linewidth]{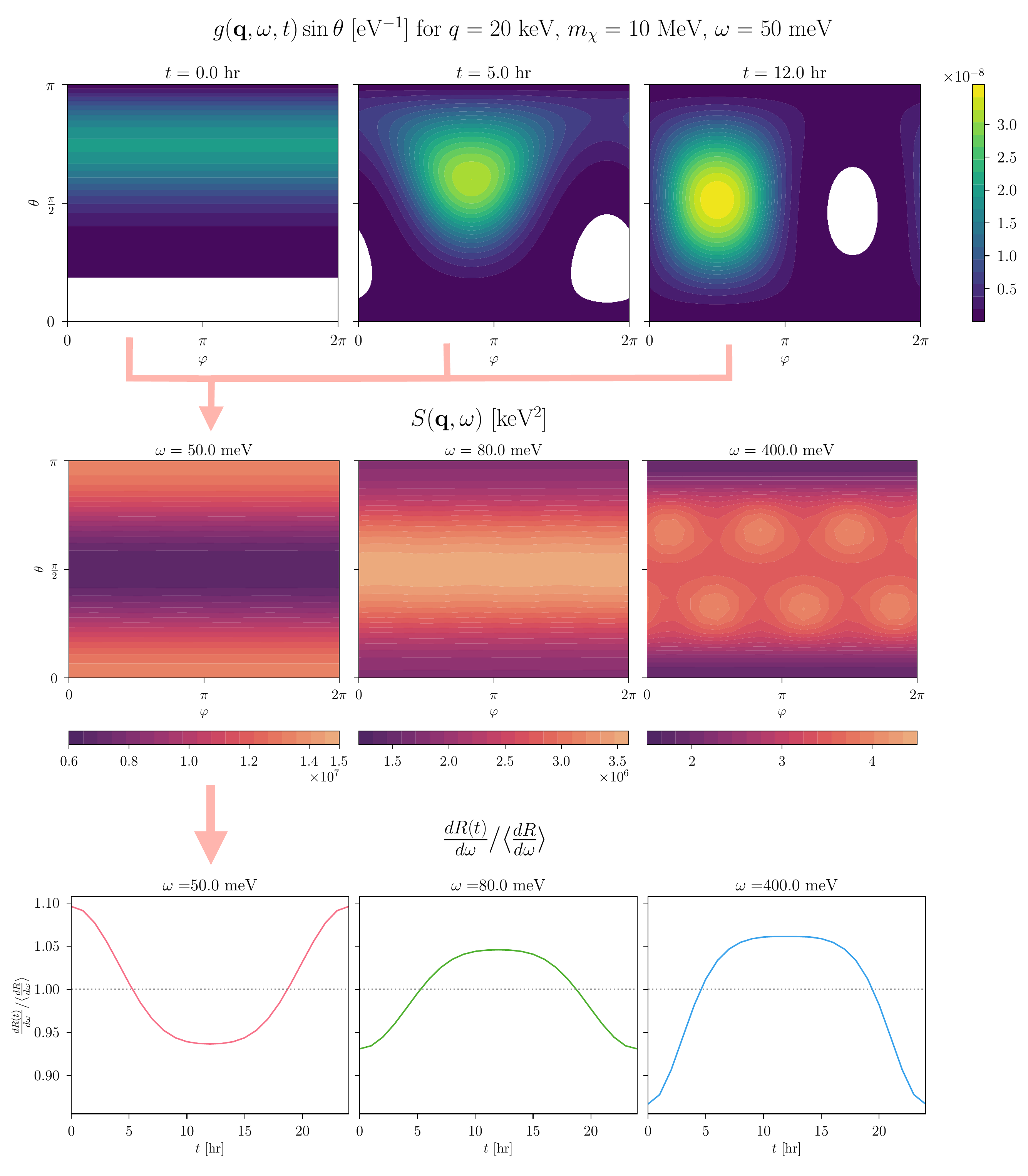}
    \caption{The kinematic function $g(\mathbf{q},\omega,t)$, weighted by $\sin{\theta}$, combines with the structure factor $S(\mathbf{q},\omega)$ to give rise to daily modulation in $\frac{dR(t)}{d \omega}$. The first row shows the kinematic function $g(\bfq,\omega,t)$ over different orientations of $\bfq$, described by $\theta$ and $\varphi$, for fixed $q = 20$ keV, $m_{\chi} = 10$ MeV, and $\omega = 50$ meV and at three times $t$. White regions are forbidden phase space where $g(\bfq,\omega,t) = 0$. The second row is the structure factor for Al$_2$O$_3$ at the same $q$, but at different $\omega$. The arrows illustrate how the kinematic function at different times $t$ combines with the structure factor to create a daily modulation in the differential rate $\frac{dR (t)}{d \omega}$ for $m_\chi = 10$ MeV (third row). 
    }
    \label{fig:g_qwt,S_qw_directional}
\end{figure*}

We turn to Fig.~\ref{fig:g_qwt,S_qw_directional} to explore the interplay between the kinematic function and the structure factor, and how they combine to create daily modulation. It is convenient to evaluate Eq.~\eqref{eq:total_rate} in spherical coordinates for $\bfq$. As standard, $\theta$ is taken as the angle of declination from the crystal's primary $\hat{z}$ axis, and $\varphi$ is a rotation around $\hat z$, with $\varphi=0$ along the crystal's $\hat x$ axis.

In the first row we show $g(\bfq,\omega,t) \sin{\theta}$  over a grid of $\theta$ and $\varphi$ at different times $t$ of the day. 
The values of $q$, $\omega$ and $m_{\chi}$ are fixed to be in the sensitive region of $v_{*}$. This choice illustrates how there is a forbidden region in $\hat q$ where $g(\bfq,\omega,t)=0$, which changes throughout the day. At any time during the day $t$, the kinematic function has preferred directions $\hat q_{\text{pref}}(t)$, denoted by the bright regions in the first row. At $t=0$, this region is concentrated in a cone at angle $\pi /4$ from the primary $\hat z$ axis of the crystal, as the axis is aligned with $\bfv_e$.

The preferred directions of $g(\bfq,\omega,t)$ depend on $q/2m_{\chi}$. For $q$ values larger than the one shown in Fig.~\ref{fig:g_qwt,S_qw_directional}, the preferred directions remain approximately the same as shown. This angular dependence illustrates the main effect for massive mediators where scattering is dominated by high $q$. For lower $q$ or higher $m_{\chi}$, a wider range of angles $\theta$ can contribute, so the $xy$ plane tends to dominate due to the $\sin{\theta}$ factor in the integral. This leads to somewhat different modulation effects for massless mediators, lower number of phonons and higher masses.

In the second row, we show $S(\bfq,\omega)$ again over a grid of $\theta$ and $\varphi$ describing the orientation of $\bfq$. We fix the same $q$ but varying $\omega$ in successive plots. At different $\omega$ values, the structure factor favors its own direction $\hat q$. This is a consequence of the anisotropies in Al$_2$O$_3$ that manifest through the pDoS, as explained in Fig.~\ref{fig:Al2O3_Sqw}.

Finally, in the third row, we demonstrate how the kinematic function combines with each plot in row two to create a daily modulation in the differential rate $\frac{d R(t)}{d \omega}$, as given by Eq.~\eqref{eq:total_rate}. 
When the structure factor is weighted in the same $\bfq$ directions as the kinematic function, the differential rate $\frac{dR(t)}{d \omega}$ will be higher. This modulation is most dramatic in the region of $\omega \sim$ 50 meV where the structure factor is weighted towards $\hat q = \pm \hat z$. The  function $g(\bfq,\omega,t)$ is weighted in a similar direction at $t=0$, but at $t=12$ hr, is weighted along $\hat q = \hat y$. Consequently, for this $\omega$, the differential rate is higher than average at $t=0$, but lower than average at $t=12$ hr. 
We see the opposite effect for high $\omega \gtrsim 100$ meV, where the structure factor is weighted towards the $xy$ axis, leading to an inverted modulation plot. It is important to notice the magnitude of the structure factor and hence the contribution to the total rate in these $\omega$ regions. At these $q$ values, the structure factor is orders of magnitude higher for $\omega \lesssim 100$ meV, while the modulation for high $\omega$ is appearing in the tail of the scattering distribution where the rate is suppressed.

\section{Results \label{sec:results}}

In this section, we present modulation results in sapphire (Al$_2$O$_3$) for both scalar and dark photon mediators, in the massless and massive limits. First, we introduce some definitions.

To see how the anisotropies manifest at different recoil energies $\omega$, we define the \textit{differential} daily modulation amplitude:
\begin{align}
    \label{eq:daily_diff_mod_amplitude}
    A_{mod}\left(\frac{dR}{d \omega} \right) \equiv \frac{\frac{dR(t=0)}{d \omega} - \langle \frac{dR}{d \omega} \rangle }{\langle \frac{dR}{d \omega} \rangle}.
\end{align}
We define this using the differential rate at $t=0$ as that is when it exhibits the greatest deviation from average, as seen in the bottom three panels of Fig.~\ref{fig:g_qwt,S_qw_directional}.

To evaluate the amount of modulation in the total rate, we define the daily modulation amplitude as 
\begin{align}
    \label{eq:daily_mod_amplitude}
    A_{mod} \equiv \max \frac{|R(t) - \langle R \rangle |}{\langle R \rangle}.
\end{align}
Once again, noting that $A_{mod}$ occurs at $t=0$, we speed up computation by calculating $A_{mod} = |\frac{R(t=0) - \langle R \rangle}{\langle R \rangle} |$.

Finally, we define a cross section $\sigma_{mod}$ and number of events $N_{ev}$, needed to see the modulation. Since the advantage of modulation is that it enables background rejection, we also account for a potential source of backgrounds with rate $R_B$. For phonon-only signals, modeling the origin of backgrounds and predicting their rate is challenging. We will therefore assume that $R_B$ is {\it a priori} unknown.

First, consider the ideal case of zero background rate $(R_B = 0)$. The non-modulating portion of the DM signal itself acts as a background for the modulation, since $R_B$ is unknown.
Assuming a roughly sinusoidal signal, we can split the day into two halves, where the modulation is above and below average respectively. Given $N_{ev}$ total events, we expect there to be $N_{ev}/2$ events in each half of the day. So, the size of our signal is $\frac{N_{ev}}{2} A_{mod}$, while the size of fluctuations for a uniform (time-independent) signal would be $\sqrt{N_{ev}/2}$. Performing a $S/ \sqrt{B}$ estimate for significance in each half of the day and adding them in quadrature, we can see that to establish a statistically significant modulation at the 2$\sigma$ level, we require
\begin{align}
    \frac{N_{ev} A_{mod}}{\sqrt{N_{ev}}} = 2 \quad \rightarrow \quad N_{ev}(R_B = 0) = \frac{4}{A_{mod}^2}.
\end{align}
For each $m_{\chi}$ and $E_{thresh}$, we can solve for the cross section $\sigma_{mod}$ needed to observe this number of events with exposure $\mathcal{E}$:
\begin{align}
    \label{eq:sigma_mod}
      \sigma_{mod} \equiv N_{ev} \frac{\sigma_0}{\langle R \rangle \cdot \mathcal{E}}
\end{align}
where $\sigma_0 = 10^{-38}\, {\rm cm}^2$ is the reference cross section we use in calculating the average rate $\langle R \rangle$. While this treatment of $\sigma_{mod}$ is highly simplistic, we note that it reproduces  well the result of more sophisticated statistical analyses in previous work~\cite{Griffin:2018bjn,Coskuner:2021qxo}. For the same exposure and in the absence of backgrounds, the cross section sensitivity of a total rate measurement is better by a factor of $\sim 1/A_{mod}^2$.

This simple treatment can now be used to estimate the impact of additional backgrounds with rate $R_{BG}$, which are constant in time. 
Our estimate of $\sigma_{mod}$ above is  modified by including $R_{BG}$ in the background fluctuations, $\sqrt{B} \rightarrow \sqrt{N_{ev} + R_{BG} \mathcal{E} }$ for exposure $\mathcal{E}$, which gives the required number of DM events
\begin{align}
    N_{ev} &= \frac{2}{A_{mod}^2} \left(1 + \sqrt{1 + A_{mod}^2 R_{BG} \mathcal{E} } \right) \\
    &\approx
    \begin{cases} 
    4/{A_{mod}^2}, & A_{mod}^2 R_{BG} \mathcal{E} \ll 1 \\
    2 \sqrt{R_{BG} \mathcal{E}}/{A_{mod}} , & A_{mod}^2 R_{BG} \mathcal{E} \gg 1 \\
    \end{cases}.
    \label{eq:Nev_RB}
\end{align}
Therefore, the number of events either is independent of $R_{BG}$, or at worst scales as $\sqrt{R_{BG}}$ in the limit of large $R_{BG}$. In contrast, since $R_{BG}$ is  unknown, the sensitivity of the total rate measurement is then limited to cross sections giving $R > R_{BG}$. For high $R_{BG}$, it is therefore possible for modulation to give {\emph{similar or better sensitivity}} than a total rate measurement. Finally, we emphasize that if $R_{BG}$ is unknown, the modulation signal offers the potential for a discovery while total rate measurements can only set upper bounds. Accounting for energy-dependence of the modulation rate would potentially also provide information about the DM model in the event of a signal.

In this work, we consider illustrative background rates of $R_{BG} = $1/gram-day, 1/gram-hour and a fixed exposure of $\mathcal{E} = \textrm{kg}\cdot\textrm{year}$. For each case, we will compare cross section sensitivity of the total rate measurement to $\sigma_{mod}$, which is again determined using Eq.~\ref{eq:sigma_mod}.

All computations are implemented in \texttt{DarkELF}. When the modulation amplitude becomes small, increased sampling of the integral is necessary. We describe our default numerical settings and convergence tests for the modulation calculation in Appendix~\ref{convergence_test_appendix}. The \texttt{DarkELF} implementation is summarized in Appendix~\ref{darkelf_appendix}. Results for SiC are shown in Appendix~\ref{additional_results_appendix}, while other materials we investigated (Si, GaAs and SiO$_2$) had negligible modulation.

\subsection{Heavy scalar mediator}

\begin{figure*}
    \centering
    \includegraphics[width=1\linewidth]{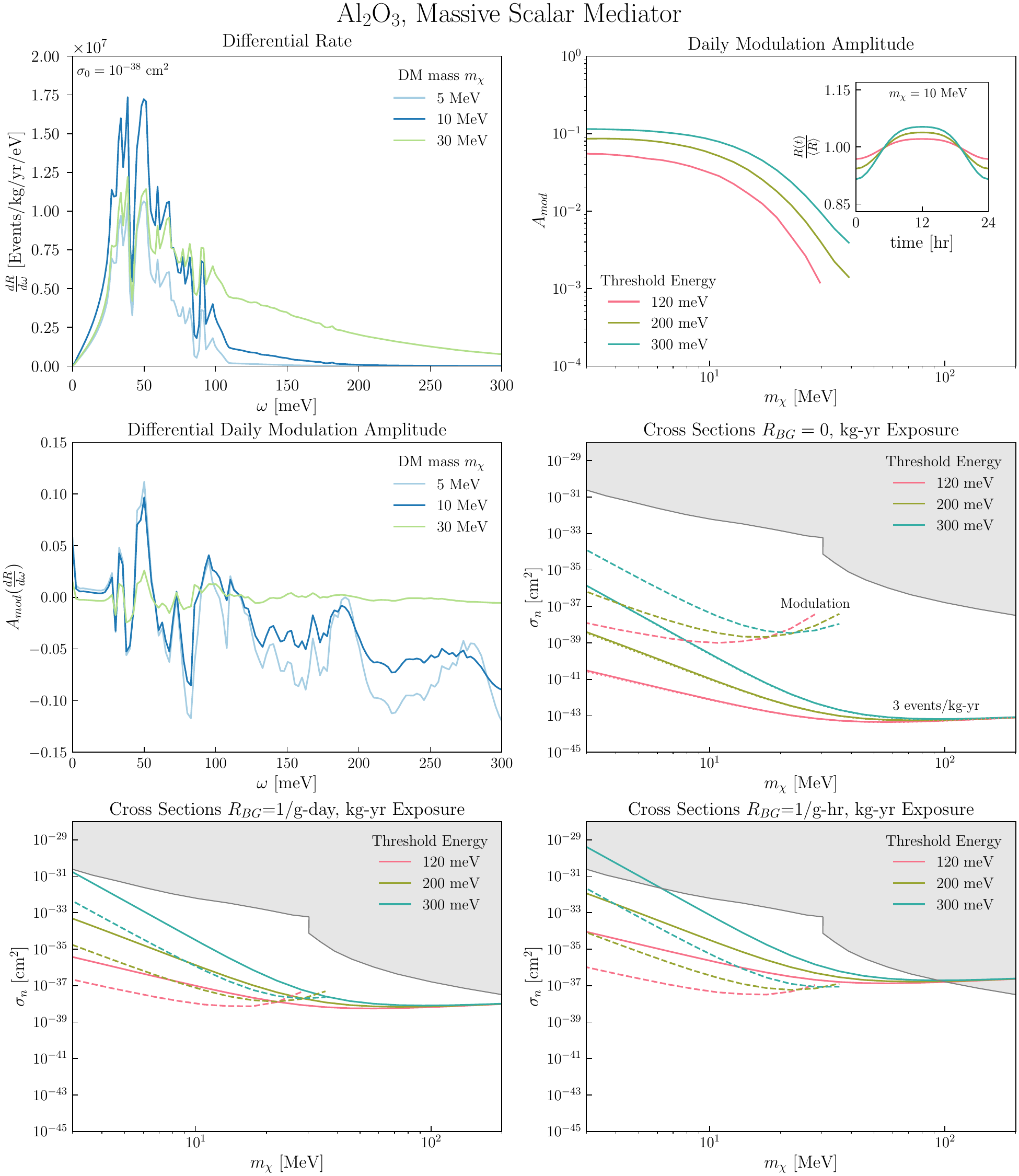}
    \caption{Results for Al$_2$O$_3$ detector, assuming a massive scalar mediator. \textbf{Top left:} Differential rate $\frac{dR}{d \omega}$ for a reference cross section of $\sigma_0 = 10^{-38}$ cm$^2$. \textbf{Top right:} Daily modulation amplitude $A_{mod}$ as a function of $m_{\chi}$, for a few different energy thresholds. Also shown is the daily modulation curve $\frac{R(t)}{\langle R \rangle}$ at $m_{\chi} = 10$ MeV. \textbf{Middle left:} Differential daily modulation amplitude $A_{mod}(\deriv{R}{\omega})$ for a few $m_\chi$ as a function of $\omega$. \textbf{Middle right:} Cross section curves $\sigma_n$. Solid lines are computed using the anisotropic pDoS and dotted lines are computed using the isotropic approximation, both assuming 3 events/kg-yr. Dashed lines are $\sigma_{mod}$, the cross section to observe a modulating signal. \textbf{Bottom left:} Similar to middle right, but with $R_{BG}=$ 1/gram-day. Here solid lines are sensitivity of a total rate measurement while dashed is sensitivity of a modulation measurement. \textbf{Bottom right:} Same as bottom left, with $R_{BG}=$ 1/gram-hr.
    }
    \label{fig:al2o3_massive_scalar}
\end{figure*}

In the case of a massive scalar mediator coupling equally to nucleons, $f_d^2 = A_d^2$ and $|\Tilde{F}(q)|^2 = 1$. We present results in Fig.~\ref{fig:al2o3_massive_scalar}. First, in the top left and middle left plots, we show the total differential rate $\frac{dR}{d \omega}$ and the differential daily modulation amplitude $A_{mod}
\left(\frac{dR}{d \omega} \right)$, for the same set of masses and the same $\omega$ range. In general, there is a strong $\omega$ dependence in the daily modulation which can be explained by the anisotropies in the structure factor at different $\omega$.

In the middle left of Fig.~\ref{fig:al2o3_massive_scalar}, the drastic sharp peaks at low $\omega$ can be explained by anisotropies in the pDoS. In particular, the peaks resemble the structure factor in Fig.~\ref{fig:Al2O3_Sqw} where the $n=1$ term directly conveys anisotropies from the pDoS. In the region $\omega \sim 50$ meV, there is greater than average rate at $t=0$, as the structure factor is aligned with the kinematic function. However, for $60 \lesssim \omega \lesssim 85$ meV, the structure factor prefers the $xy$ plane, resulting in the opposite phase of modulation at $t=0$. Note that the $m_{\chi}=50$ MeV line has the opposite phase compared to the lower masses. This is a result of the kinematic function shifting its preferred direction at $t=0$ towards the $xy$ plane for lower values of $q/2m_{\chi}$.

We can also explain the oscillations at high $\omega$ as a feature of the tails of individual $n$-phonon peaks in $S(\bfq,\omega)$. At the low $q$ relevant for the DM masses, for any range of $\omega$, scattering is dominated by a particular $n$. Within the range of $\omega$ where a given term dominates, the pattern of modulation favors the $\hat z$ direction at lower energies and $xy$ plane at higher energies, as discussed in Fig.~\ref{fig:Al2O3_Sqw_nphonon}. As $\omega$ increases further and the dominant phonon term shifts to $n+1$, the structure factor starts to shift back towards $\hat z$ again. For example, the $n=4$ term in Figure \ref{fig:Al2O3_Sqw_nphonon} switches from being weighted along $\hat z$ to $xy$ at $\omega \sim 230$ meV. This corresponds exactly with where there is a local minimum in modulation in the middle left plot of Fig~\ref{fig:al2o3_massive_scalar}.

In general, there is an increased modulation for large $\omega$. This is a result of the weighting of $S(\bfq,\omega)$ towards the $xy$ plane at high $\omega$, as can be seen in both Figs.~\ref{fig:Al2O3_Sqw}-\ref{fig:g_qwt,S_qw_directional}. With increasing $\omega$, the differential rate modulation curves all resemble the bottom right panel of Fig.~\ref{fig:g_qwt,S_qw_directional} as the structure factor strongly aligns with the kinematic function's preferred direction $\hat q_{\text{pref}}(t)$ at $t=12$ hr. It is important to note that despite this increased modulation at higher energies, the total rate is highly suppressed in this region, as evident in the top left panel of Fig.~\ref{fig:al2o3_massive_scalar}.

Next, we see how modulation as a function of $\omega$ results in modulation in the total integrated rate. In the upper right panel, we show the daily modulation amplitude $A_{mod}$ as a function of $m_{\chi}$. The inset shows daily rate modulation $\frac{R(t)}{\langle R \rangle}$ over a sidereal day for a DM mass $m_{\chi}=10.0$ MeV at a few values of $E_{thresh}$. We find that Al$_2$O$_3$ has a daily modulation amplitude up to 11\% for a DM mass $m_{\chi} = 10$ MeV and $E_{thresh} = 300$ meV. We only show energy thresholds above the single phonon regime, $\omega \geq 120$ meV, and for these energies, as seen in the middle left panel, the differential modulation is peaked at the same time of day so the total modulation becomes more pronounced.

Overall, the most modulation occurs for $m_{\chi} \lesssim 30$ MeV, above which modulation subsides. This matches our intuition of an upper mass limit on modulation. Greater mass corresponds to greater momentum transfers $q$, and as seen in Figure \ref{fig:Al2O3_Sqw}, the effects of anisotropy subside at large $q$ upon approaching the nuclear recoil limit. Note that we have limited our maximum mass value to 30-40 MeV, depending on the energy threshold. At higher masses, the modulation becomes very small and requires many more sampling points to calculate accurately; see Appendix \ref{convergence_test_appendix} for a more detailed description of the convergence tests for the modulation calculation.

In the middle right panel we show cross section curves for a rate of 3 events/kg-yr assuming zero background, computed using the full $\bfq$ integral (solid lines) or the isotropic approximation (dotted lines).  The isotropic approximation turns out to be a good estimate for $m_{\chi} \gtrsim 1$ MeV, even for an anisotropic material. For $m_{\chi} \lesssim 1$ MeV, single phonon production becomes the dominant scattering mechanism. These have already been computed numerically, including modulation amplitudes, in Refs.~\cite{Griffin:2018bjn,Coskuner:2021qxo}. We exclude single phonon contributions to the total rate in our calculations and institute a lower cutoff on the $q$-integral of $q\geq q_{BZ} = \frac{2 \pi}{a}$ for $a$ the lattice constant. In sapphire, we take the cutoff as $q_{BZ}=0.95$ keV. This leads us to set the lower limit of the curves at $m_{\chi} = 3$ MeV.

We also show the cross sections $\sigma_{mod}$ for detecting a modulating signal at the same energy thresholds assuming zero background. As expected, modulation is most effective for direct detection at low $m_{\chi}$. Note that the masses at which the modulation amplitude is largest corresponds to the masses and thresholds where the total rate decreases significantly. So while higher energy thresholds may exhibit greater modulation, they come with a significantly decreased overall rate.

In the last row, we show the cross sections corresponding to detecting both a total rate measurement (solid) and a modulating signal (dashed) for a background rate of $R_{BG}=$ 1/gram-day and 1/gram-hour. Notably, modulation can offer improved or comparable reach compared to total rate detection for high background rates. In particular, the cross section reach for modulation has barely shifted even when $R_{BG} = 1/$g-hr. This is because the modulation fraction is small, putting us in the regime $A_{mod}^2 R_{BG} \mathcal{E} \ll 1$ of Eq.~\ref{eq:Nev_RB}, so the values of $R_{BG}$ do not significantly affect the significance of the modulating signal.

In the cross section plots, existing direct detection constraints are shown in shaded gray. These include indirect probes of nuclear recoils via the Migdal effect with PandaX-4T~\cite{PandaX:2023xgl}, DarkSide-50~\cite{DarkSide:2022dhx}, XENON1T~\cite{XENON:2019zpr} as well as SENSEI@SNOLAB~\cite{SENSEI:2023zdf}.  In addition to direct constraints, there are model-dependent upper bounds on the possible achievable cross sections, for a discussion see for example Refs.~\cite{Knapen:2017xzo,Elor:2021swj,Bhattiprolu:2022sdd}.

Note that the existing constraints do not extend to arbitrarily high cross sections due to attenuation of DM in the Earth's crust or atmosphere. We have cut off our plots at cross sections of $10^{-28}$~cm$^2$, roughly where DM scattering in the Earth crust would significantly impact direct detection in underground labs~\cite{Emken:2019tni}.

\subsection{Massless scalar mediator}
\label{sec:massless_scalar}

\begin{figure*}
    \centering
    \includegraphics[width=1\linewidth]{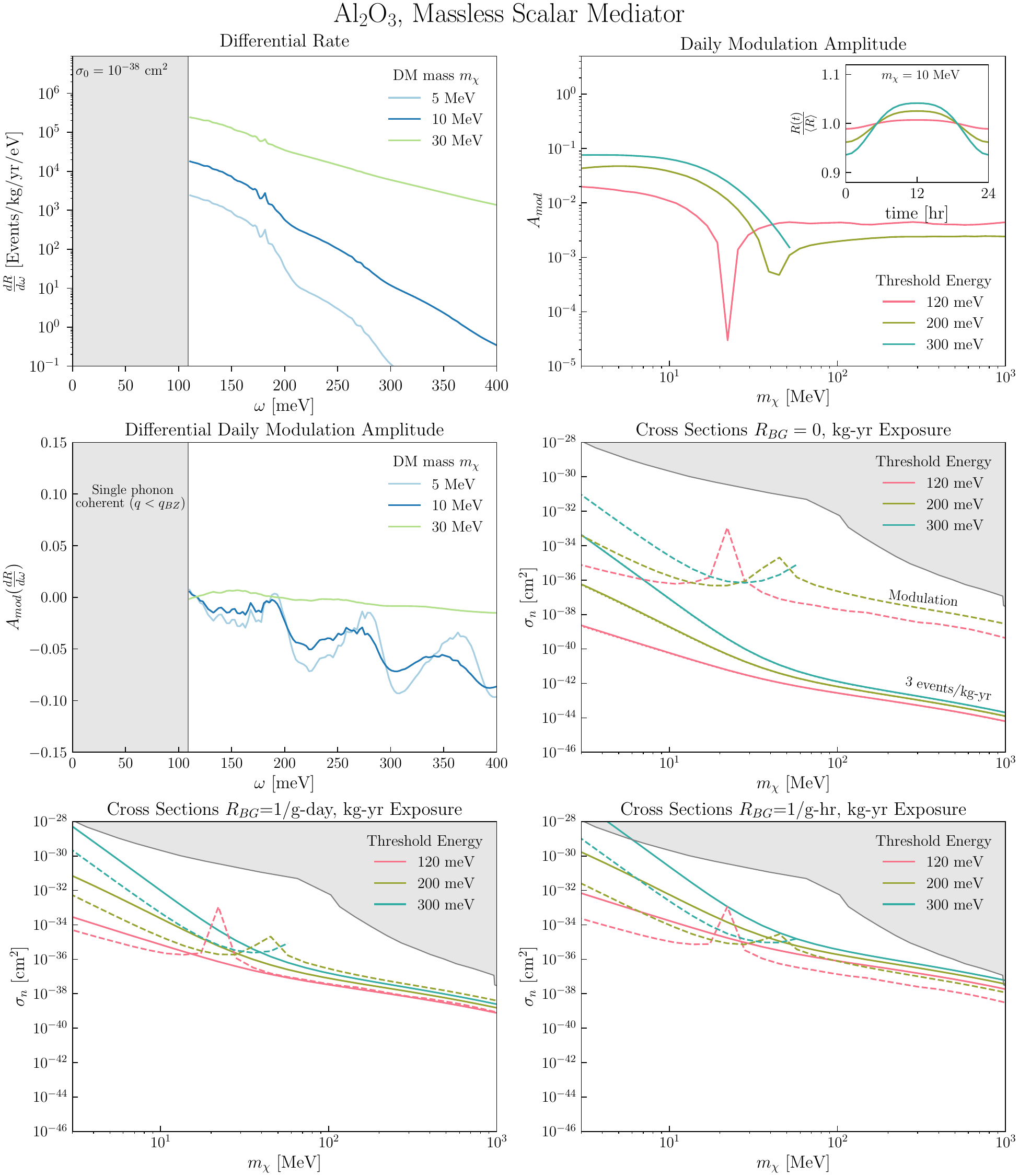}
    \caption{Same as Figure~\ref{fig:al2o3_massive_scalar}, for the massless scalar mediator. The regime where the coherent single phonon term is expected to dominate is shaded gray in the upper left and middle left plots. 
    }
    \label{fig:al2o3_massless_scalar}
\end{figure*}

For a massless scalar mediator, we now take $f_d^2 = A_d^2$ and $|\Tilde{F}(q)|^2 = (\frac{m_{\chi} v_0}{q})^4$ with $v_0 = 220$ km/s and present results in Figure~\ref{fig:al2o3_massless_scalar}. First, we briefly review the effect of the form factor on the total rate; see Ref.~\cite{Campbell-Deem:2022fqm} for a more complete description. The two primary adjustments to the integrand are the additional $m_{\chi}^4$ and $1/q^4$ scaling. The weighting towards high masses leads to increased rates and cross section curves which improve with higher masses. As for the additional $1/q^4$ scaling, for $n=1$ the integrand is weighted towards the lowest $q$ and the single phonon coherent piece dominates. We therefore exclude single phonon excitations (shaded region in plots), see Ref.~\cite{Griffin:2018bjn,Coskuner:2021qxo} for results in that regime. For $n \ge 2$, the integrand is still weighted at high $q$ such that our calculations apply.

Now, we turn our attention to the effect on modulation. We again see similar modulation at high $\omega \gtrsim 100$ meV as the massive scalar mediator in the lower left plot. This results in similar, although slightly decreased total modulation to the massive mediator, with only 7.6\% modulation for $E_{thresh} = 300$ meV and $m_{\chi} = 10$ MeV. Overall, the daily modulation amplitude as a function of mass looks relatively similar to the massive mediator case, although in this case they tend to plateau to a constant (but very small) value with higher $m_\chi$ rather than going to zero. For the massless mediator, we include results for a larger range of mass values than the massive mediator, according to the results of our convergence tests in App.~\ref{convergence_test_appendix}.

For the cross section curves, we observe similar results for the modulation curves $\sigma_{mod}$ relative to $\sigma_n$ as for the massive mediator. We again see masses where $\sigma_{mod}$ appears to spike, due to  the total modulation amplitude going to zero at those masses. However, the differential modulation is still nonzero and with good energy resolution, it is possible to search for modulation in different parts of the energy spectrum. Compared to the sensitivity of a total rate measurement, we again see a similar or improved reach for a modulating signal when considering a non-zero background.

As for the heavy scalar mediator, we show existing constraints (gray) from probes of nuclear recoils with the Migdal effect with XENON1T~\cite{XENON:2019zpr} as well as SENSEI@SNOLAB~\cite{SENSEI:2023zdf}. Model-dependent upper bounds on achievable cross sections are discussed in Ref.~\cite{Knapen:2017xzo}.

\subsection{Massive dark photon}

\begin{figure*}
    \centering
    \includegraphics[width=1\linewidth]{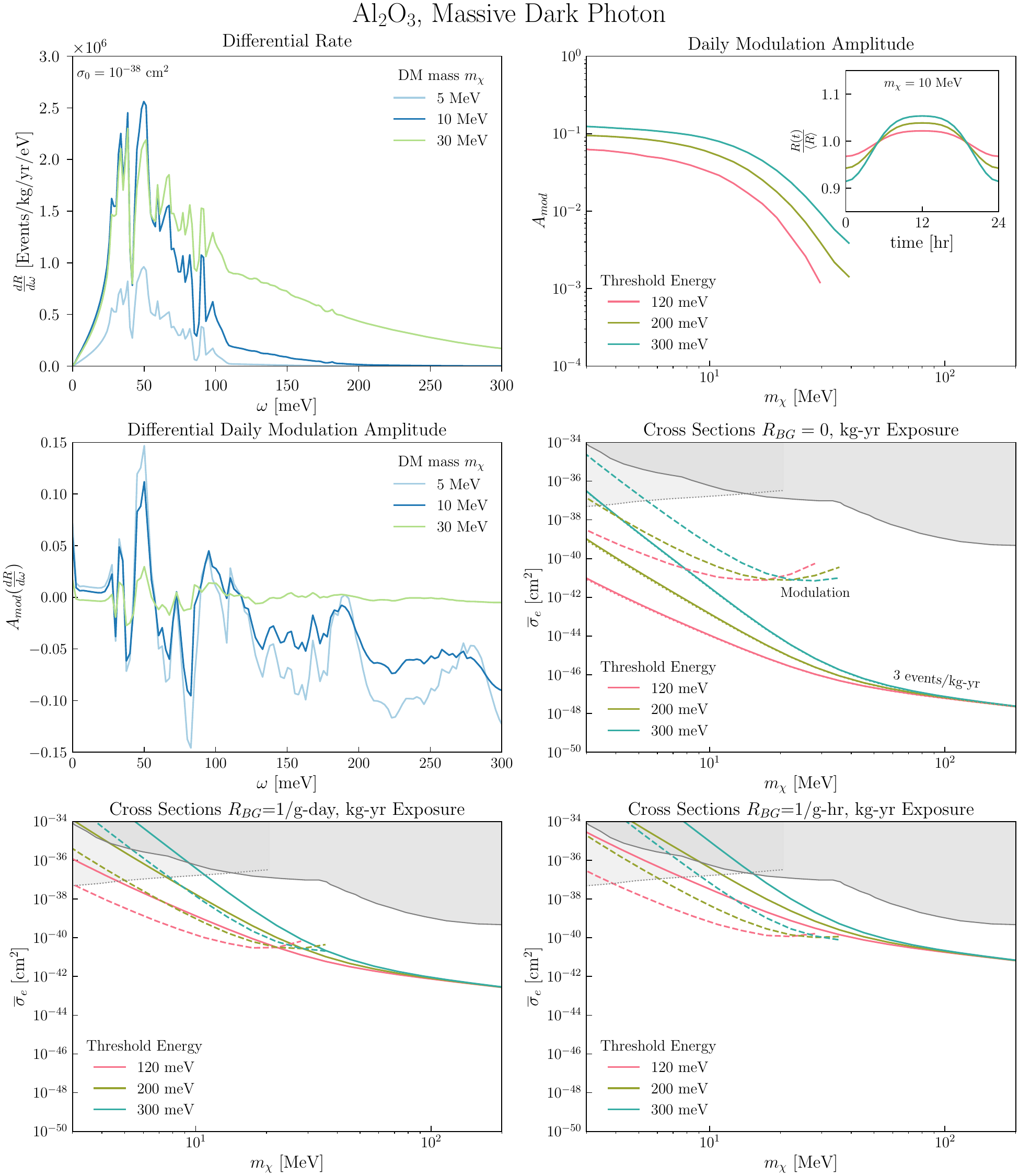}
    \caption{Same as Figure~\ref{fig:al2o3_massive_scalar}, for the massive dark photon mediator. Note that we now show the DM-electron cross section $\overline \sigma_e$ in the cross section plots. Additionally, the shaded gray are exclusions from current experiments, including constraints on halo DM (solid) with SENSEI@SNOLAB~\cite{SENSEI:2023zdf}, DAMIC-M~\cite{DAMIC-M:2023gxo}, DarkSide-50~\cite{DarkSide:2022knj}, and XENON1T~\cite{Aprile:2019xxb}; the dotted shows constraints on solar reflected dark matter~\cite{emken2024solarreflectiondarkmatter}.
    }
    \label{fig:al2o3_massive_darkphoton}
\end{figure*}

A dark photon mediator couples to the electric charge of a SM particle. In the regime where phonons are relevant, the charge of the nucleus is partially screened by the electrons. So, we use the notion of a momentum dependent $\textit{effective charge}$ seen by the DM. Using the calculations in~\cite{BrownXray}, we take $f_d = Z_d(q)$ with $Z_d(q)$ the effective charge for each atom $d$. The effective charge interpolates between 0 at low $q$ to the nuclear charge at high $q$. The approximation is expected to be valid for $q \gtrsim q_{BZ}$, in accordance with the rest of our multiphonon calculations. For the massive dark photon, we again take $|\Tilde{F}(q)|^2 = 1$.

The results for a massive dark photon mediator, shown in Figure~\ref{fig:al2o3_massive_darkphoton}, are very similar to the massive scalar mediator. The differential rate plots are diminished by a factor of $\sim 5-10$ for the massive dark photon, mainly from the factor of 4 difference between the nucleon number $A_d^2$ and high momentum nuclear charge $Z_d^2$, and from the slight suppression of low $q$ in the integration. 
Once again, for the daily modulation amplitude, we see modulation up to $10\%$ for an energy threshold of $E_{thresh}=300$ meV and $m_{\chi} = 10$ MeV. The differential modulation amplitude plots are virtually identical, which makes sense as the rate integrand is identical to within an order of magnitude for $q \gtrsim 10$ keV. For lower $q$, the factor $Z_d(q)^2$  suppresses the rate, but as discussed in Fig~\ref{fig:Al2O3_Sqw}, the effects of anisotropy are present for $q \lesssim 2\sqrt{2 m_d \omega_0} \approx 95$ keV in Al$_2$O$_3$, so the screening at low $q$ does not significantly inhibit the effects of anisotropy.

Finally, for the cross section plot, we present the results in terms of the effective DM-electron cross section $\overline{\sigma}_e$ ~\cite{Essig:2015cda}, with
\begin{align}
    \overline{\sigma}_e = \frac{\mu_{\chi e}^2}{\mu_{\chi}^2}\sigma_n
\end{align}
for $\mu_{\chi e}$ the DM-electron reduced mass. Overall, results are similar as for the massive scalar mediator, aside from the effect of the prefactor $\mu_{\chi e}^2/\mu_{\chi}^2$. In the mass range that we are considering, the prefactor reduces to $m_e^2/m_{\chi}^2$, which explains the increase in rate at high mass.

The shaded gray region indicates the combined exclusion curve on heavy dark photon mediators. The solid line contains combined constraints on halo DM from SENSEI@SNOLAB~\cite{SENSEI:2023zdf}, DAMIC-M~\cite{DAMIC-M:2023gxo}, DarkSide-50~\cite{DarkSide:2022knj}, and XENON1T~\cite{Aprile:2019xxb} while the dotted line contains constraints from solar reflected dark matter~\cite{emken2024solarreflectiondarkmatter} in XENON1T~\cite{Aprile:2019xxb} (see also Ref.~\cite{An:2021qdl}). For a heavy dark photon mediator, the $\sigma_{mod}$ line is 3-4 orders of magnitude below constraints across all masses for energy thresholds $40 \text{ meV} \lesssim E_{thresh} \lesssim 200$ meV and zero background. Even with high background, the cross sections for a modulating signal are relatively similar while the cross sections for total rate detection rise considerably.

\subsection{Massless dark photon}

\begin{figure*}
    \centering
    \includegraphics[width=1\linewidth]{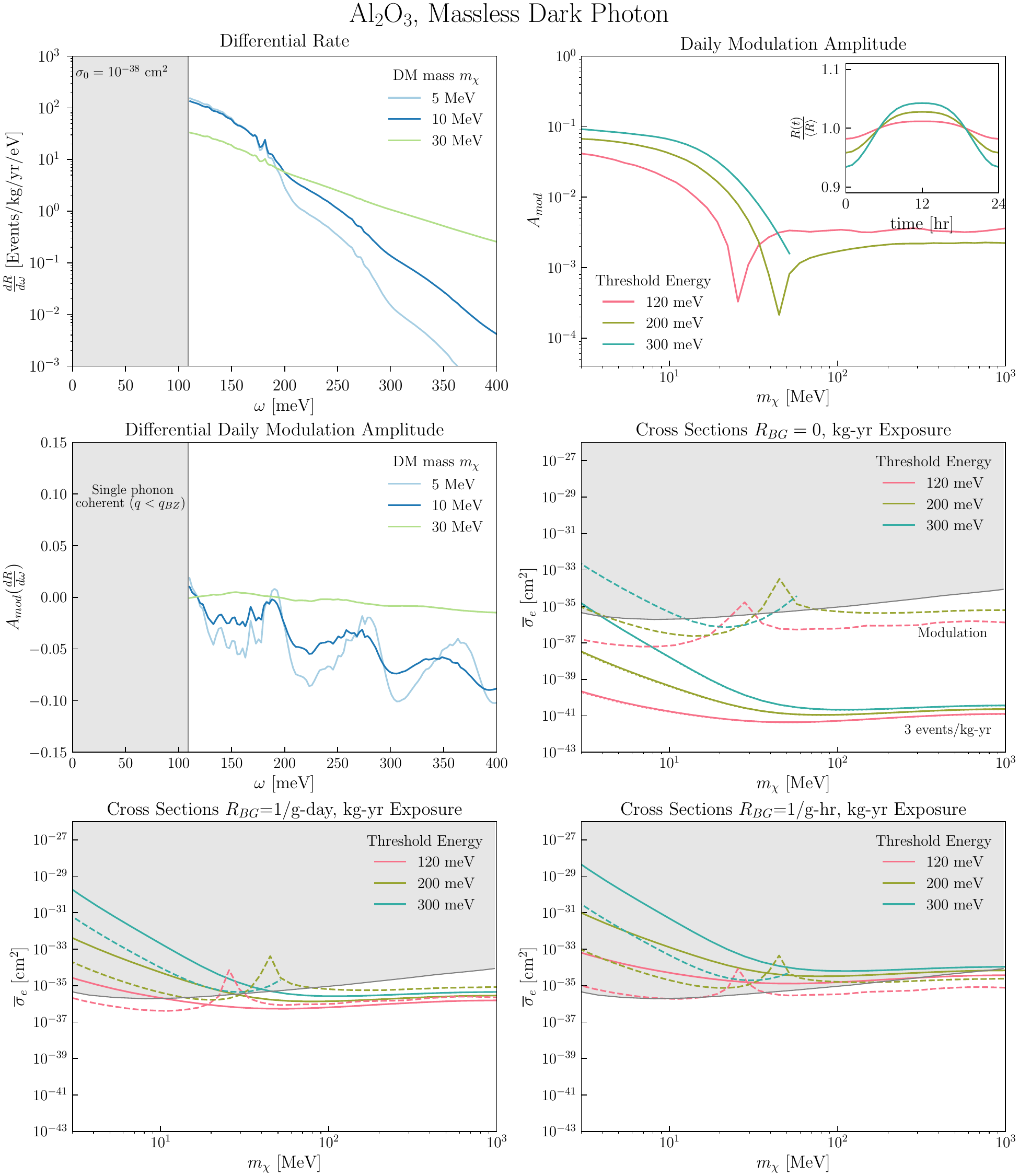}
    \caption{Same as Figure~\ref{fig:al2o3_massive_scalar}, for the massless dark photon mediator. Again, in the top left and middle left plots, the single phonon coherent regime is shaded gray. Cross section results are again shown in terms of DM-electron cross section $\overline{\sigma}_e$. The shaded region shows current constraints from  SENSEI@SNOLAB \cite{SENSEI:2023zdf}.} 
    \label{fig:al2o3_massless_darkphoton}
\end{figure*}

For a massless dark photon mediator, we use the momentum dependent effective charge $f_d^2 = Z_d^2(q)$ and the form factor $|\Tilde{F}(q)|^2 = (\frac{\alpha m_e}{q})^4$, for the reference momentum $q_0 = \alpha m_e$ where $\alpha$ is the fine structure constant and $m_e$ is the electron mass. Results are presented in Fig.~\ref{fig:al2o3_massless_darkphoton}.

Compared to the massless scalar mediator, there is a suppression at low $q$ from the $Z_d(q)^2$ term. This partially removes the preference for $q$ due to the form factor. Like the massless scalar mediator, we shade the single phonon regime to emphasize that we neglect this region of $\omega$ in calculating the total rate. In terms of the total rate, we see similar results to the massless scalar mediator. There is up to $9\%$ modulation for $E_{thresh}$ = 300 meV and $m_{\chi} = 10$ MeV.

Finally, the relation between $\sigma_{mod}$ to $\overline \sigma_e$, the cross section sensitivity for the total rate, is similar to the other cases. However, as a result of the constraints on DM-electron scattering, it is much more challenging to see modulation in phonons, as compared to the massive dark photon mediator. This is in part due to the $q$-dependence: in the massive dark photon mediator, high $q$ is preferred and DM-phonon scattering enjoys a larger $Z^2$ enhancement over DM-electron scattering. However, for the massless mediator there is a greater weight on lower $q$ where the effective change $Z_d(q)$ is more suppressed.

\section{Conclusions}

Experimental efforts to detect low energy nuclear recoils from sub-GeV dark matter will soon reach the multiphonon regime. Here DM scattering in crystals deviates from elastic nuclear recoils and results multiphonon excitations. These effects are most relevant for DM in mass range 1 MeV -- 1 GeV.  While calculations have been performed for multiphonon scattering in this mass range \cite{Campbell-Deem:2022fqm}, they have relied on the isotropic approximation. This is appropriate for isotropic crystals like Si, but it neglects important directional variation in anisotropic crystals which can lead to a daily modulation in the total scattering rate. Previously, results for daily modulation have been obtained for single phonon scattering. In this paper, we showed that daily modulation is also possible with multiphonon excitations, providing a new avenue for directional detection of dark matter.

We focused on a sapphire (Al$_2$O$_3$) detector, as a promising polar crystal target proposed for direct detection experiments.  We found appreciable modulation fractions are possible, up to 11\% for a DM mass $m_{\chi} = 10$ MeV and energy threshold $E_{thresh}=300$ meV, with decreasing modulation for higher masses and lower energy thresholds. In addition, we found similar modulation in SiC, which also has anisotropic crystal structure, but significantly less in Si, GaAs and SiO$_2$.

Even when the modulation fraction is small, it gives a powerful new method for background rejection. When backgrounds are large and {\emph{a priori}} unknown, total rate measurements are limited by the  background rate and can only provide upper bounds on DM cross sections. Searching for modulation in anisotropic crystals can give similar or better cross section sensitivity compared to total rate measurements, and has the potential to cleanly identify a signal from dark matter in our Galaxy.

\acknowledgments
We are very grateful to Simon Knapen for feedback on our draft and to Jeter Hall for the suggestion to compare cross section sensitivities for modulation and total rate measurements in the presence of a background.  CS was supported by the UCSD Undergraduate Summer Research Award and TL was supported by Department of Energy grant DE-SC0022104.

\bibliographystyle{apsrev4-1}
\bibliography{phonons}

\clearpage

\appendix

\section{Convergence Tests for Modulation} \label{convergence_test_appendix}

\begin{figure*}
    \centering
    \includegraphics[width=1\linewidth]{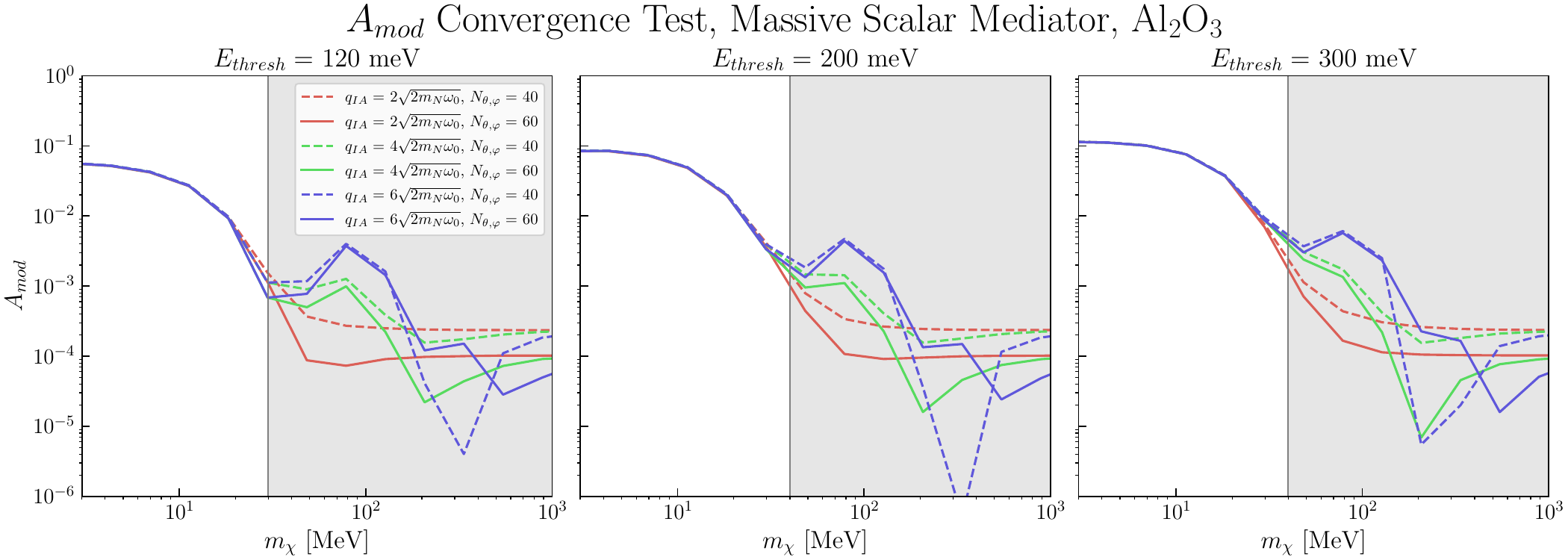}\\
    \vspace{1cm}
    \includegraphics[width=1\linewidth]{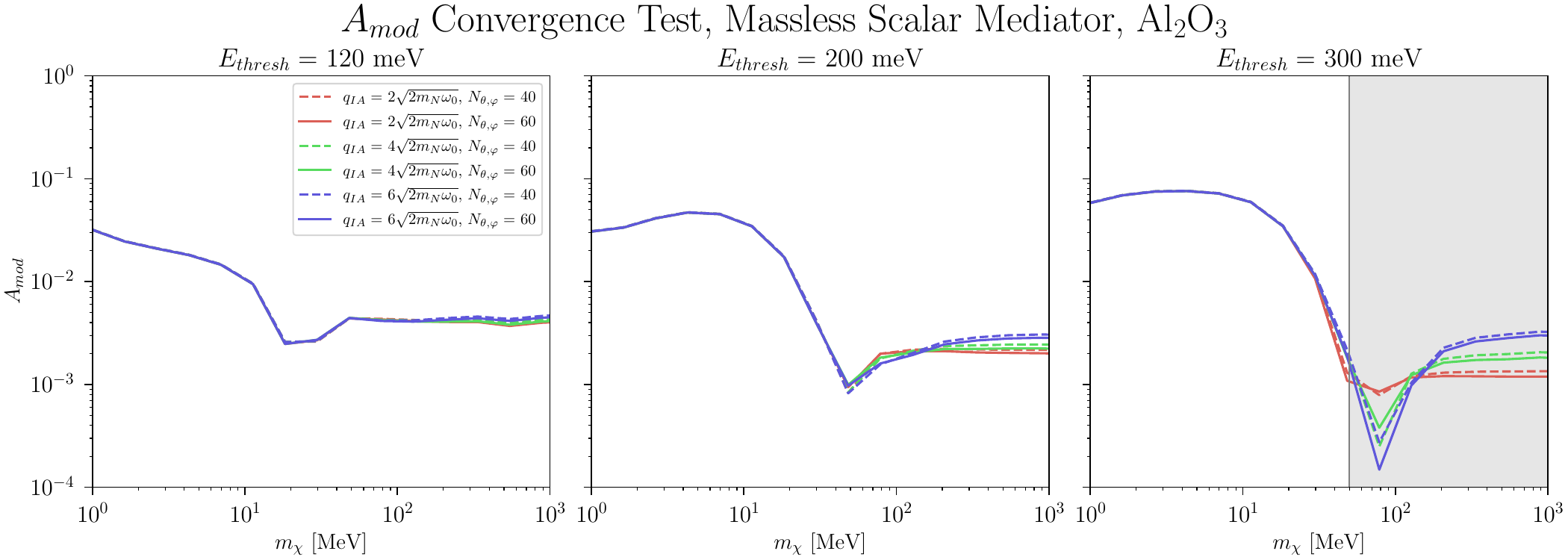}
    \caption{Convergence tests for $A_{mod}$ in Al$_2$O$_3$, for a massive scalar mediator (top) and a massless scalar mediator (bottom).}
    \label{fig:A_mod_convergence_test_sapphire}
\end{figure*}

Daily modulation is expected to decrease with higher masses, in the nuclear recoil limit. As modulation decreases, there are increasing requirements on the numerical integration for accurate calculations. Here we test the convergence of our calculations  for $A_{mod}$ with respect to the number of angular sampling points and the $q$ cutoff used for the impulse approximation. In Fig.~\ref{fig:A_mod_convergence_test_sapphire}, we show this convergence test in Al$_2$O$_3$ for massive and massless scalar mediators, respectively. 

We computed $A_{mod}$ as a function of $m_\chi$  for different combinations of $N_{\theta,\varphi}=40,60$ grid points in the angular integral and $q_{IA}=2 \sqrt{2 m_N \omega_0}, 4 \sqrt{2 m_N \omega_0}, 6 \sqrt{2 m_N \omega_0}$ for the impulse approximation. For each $m_\chi$, we compute the coefficient of variation as the standard deviation divided by the average across the different combinations, $\sigma_{A_{mod}}/\mu_{A_{mod}}$. We then restrict our results to masses where  $\sigma_{A_{mod}}/\mu_{A_{mod}} \leq 0.25$. This leads us to implement a threshold dependent upper mass limit for each mediator. For massive mediators, we restrict to $m_{\chi} \leq 30$ MeV for $E_{thresh}=120$ meV and 200 meV. For $E_{thresh}=300$ meV we restrict to $m_{\chi} \leq 40$ MeV. For massless mediators, we only restrict the $E_{thresh}=300$ meV case to $m_{\chi} \leq 50$ MeV. We use the same settings for dark photon mediators, since the results are very similar to scalar mediators. For results presented in the text, we use $q_{IA} = 4 \sqrt{m_N \omega_0}$ and $N_{\theta,\varphi}=40$ sampling points.

\section{DarkELF Implementation} \label{darkelf_appendix}

In this section, we summarize additional functions added to the \texttt{DarkELF} package~\cite{Knapen:2021bwg} that are used in the anisotropic generalization to multiphonon rate calculations. We also describe some steps for a user to perform their own computations.

We have included the necessary files to perform calculations for Al$_2$O$_3$, SiC, SiO$_2$, Si and GaAs. The user can input their own partial density of states data, but should ensure the proper formatting, listing pDoS along 6 directions in the following order: $\hat{x},\hat{y},\hat{z},\hat{x} + \hat{y}, \hat{x}+\hat{z},\hat{y}+\hat{z}$, then repeating over non-equivalent atoms. This is also described in the example pDoS files for any of the materials listed. Then, \texttt{DarkELF} will compute the components of the tensor $D_d^{ij}$. Note that in \texttt{DarkELF}, we use Eq.~\eqref{eq:T_n(omega)} to define the function $F_n^{\hat q} (\omega) = \frac{T_n^{\hat q}(\omega)}{n !}$. It is necessary to tablulate this function for each atom, which is done using \texttt{F\_n\_d\_precompute\_anisotropic}. While this is pre-tabulated for the aforementioned atoms, it must be updated for user-supplied pDoS. $\texttt{DarkELF}$ will save these tables, so the computations only needs to performed once. Also note that the mediator mass and choice of coupling are set in the same way as the isotropic calculation~\cite{Campbell-Deem:2022fqm}. Next we briefly describe the important functions.

\texttt{R\_multiphonons\_anisotropic}: \quad This function takes the time of day in hours, the energy threshold and the DM-nucleon cross section to calculate the total integrated rate in Eq~$\eqref{eq:total_rate}$. The integration over $\bfq$ and $\omega$ is performed in spherical coordinates in the following order: $q \to \theta \to \varphi \to \omega$.

\texttt{sigma\_multiphonons\_anisotropic}: \quad This function takes the energy threshold and time as inputs and returns the necessary DM-nucleon cross section to produce 3 events/kg-year. Note that this excludes the contributions from single phonons below $q_{BZ}$ (scattering in the coherent regime).

\texttt{sigma\_modulation\_anisotropic}: \quad This function takes the energy threshold as an input and returns the necessary DM-nucleon cross section to observe a modulation signal, as described in Eq~$\eqref{eq:sigma_mod}$.

\texttt{\_dR\_domega\_anisotropic}: \quad This function takes the energy transfer $\omega$, time of day $t$ and DM-nucleon cross section and returns the differential rate $\frac{dR}{d\omega}$ at that energy.

\begin{figure}[!hb]
    \centering
    \includegraphics[width=1\linewidth]{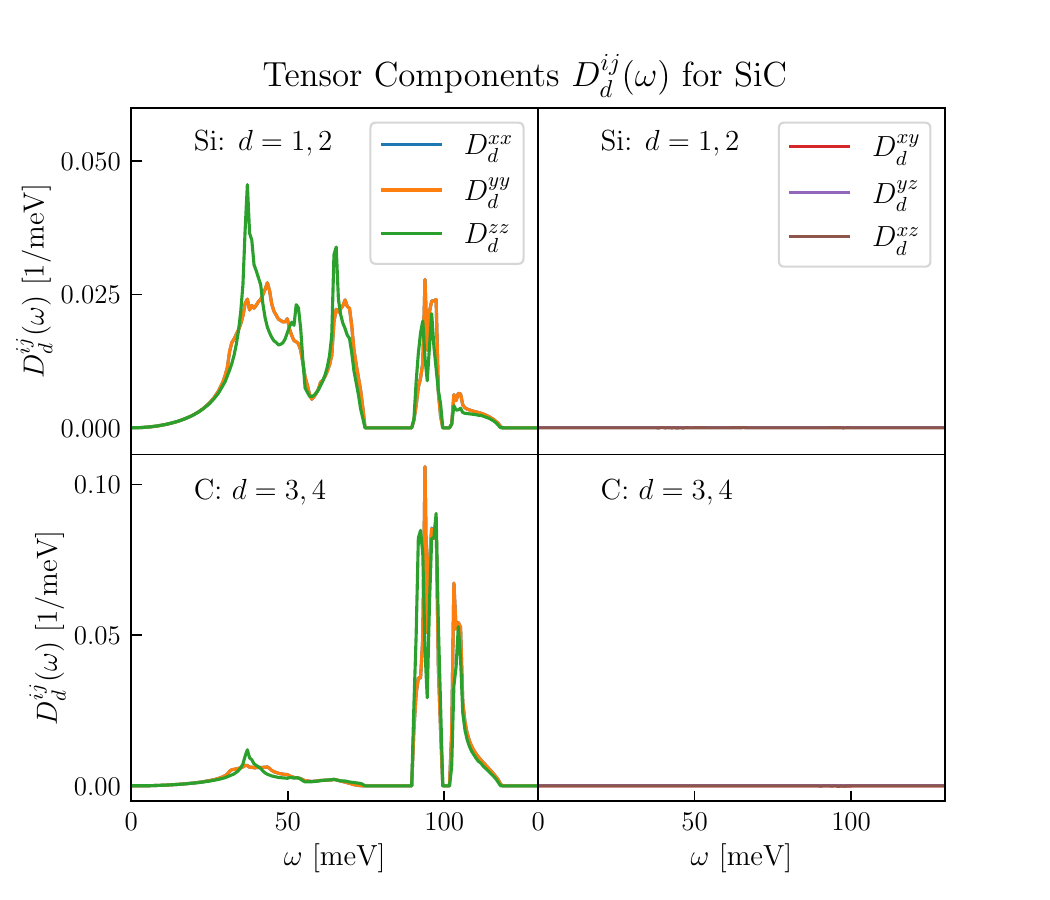}
    \caption{Density of states tensor for the 2H polytype of SiC, where redundant atoms with identical density of states are indicated. The diagonal and off-diagonal components are shown in the left and right column, respectively. Note that $D_d^{xx}$ and $D_d^{yy}$ are equivalent.}
    \label{fig:sic_pDoS}
\end{figure}

\section{Additional Results} \label{additional_results_appendix}

Here, we provide additional results for the 2H polytype of SiC, based on DFT calculations from Ref.~\cite{Griffin:2020lgd} to generate the pDoS. We selected this polytype as this structure gave the largest modulation in single phonon excitations in Ref.~\cite{Griffin:2020lgd}. 
We found negligible modulation in GaAs, Si, and SiO$_2$, so plots are not shown for these materials.

\begin{figure*}[bh]
    \centering
    \includegraphics[width=1\linewidth]{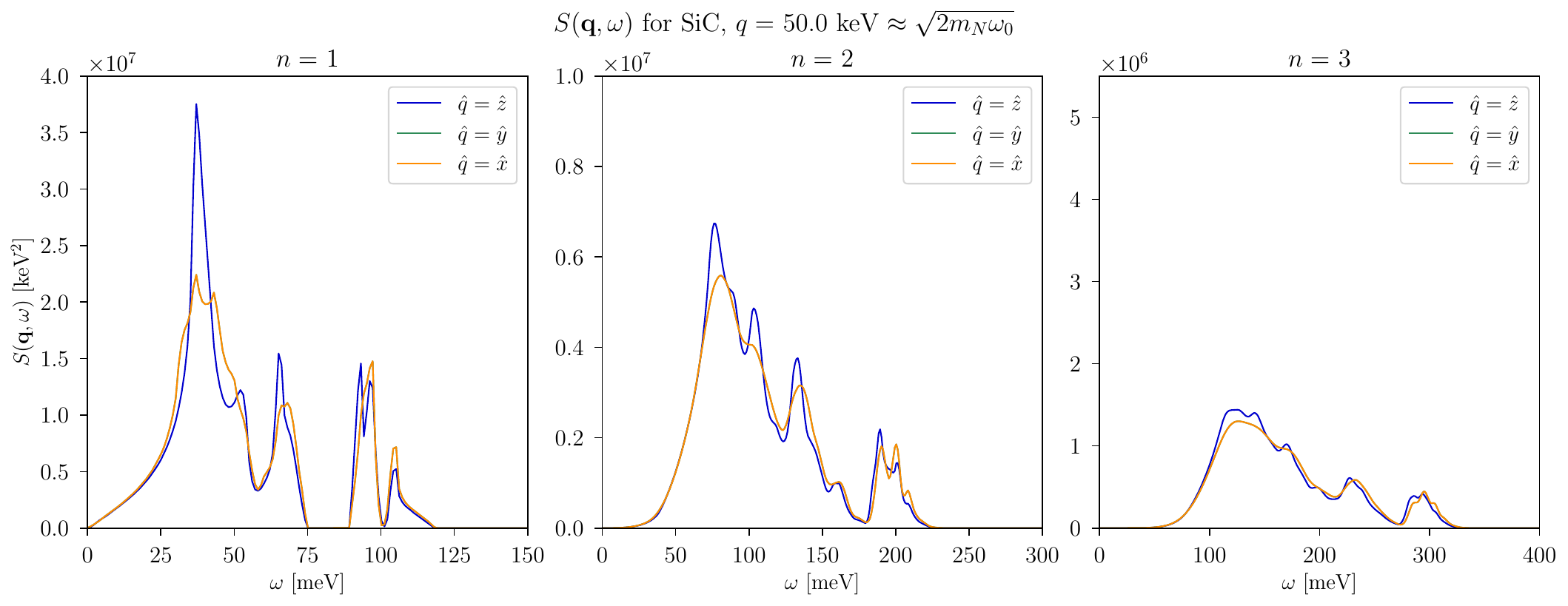}
    \caption{Decomposition of the anisotropic structure factor into $n=1,2,3$ phonon terms, for SiC and at fixed $q = 50$ keV. The structure factor is computed assuming $f_d = A_d$.  We show results for $\hat{q}= \hat x, \hat y, \hat z$. Like sapphire, the lines for $\hat x$ and $\hat y$ are identical, implying rotational symmetry with respect to $\hat z$. 
    Note that the axis scales are different in all panels. }
    \label{fig:sic_S_qw}
\end{figure*}

\begin{figure*}
    \centering
    \includegraphics[width=1\linewidth]{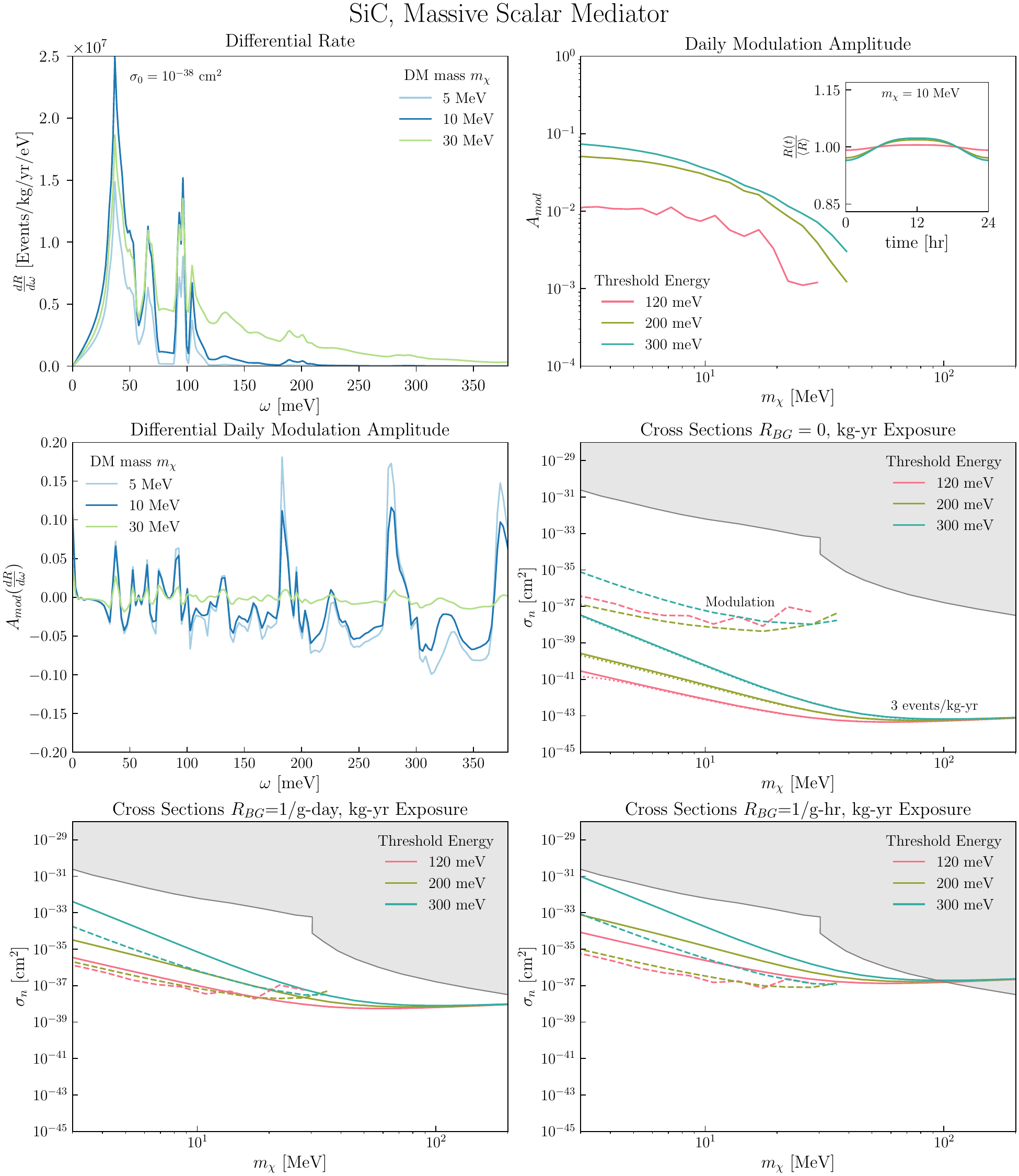}
    \caption{Results for SiC detector and massive scalar mediator.}
    \label{fig:sic_massive_scalar}
\end{figure*}

\begin{figure*}
    \centering
    \includegraphics[width=1\linewidth]{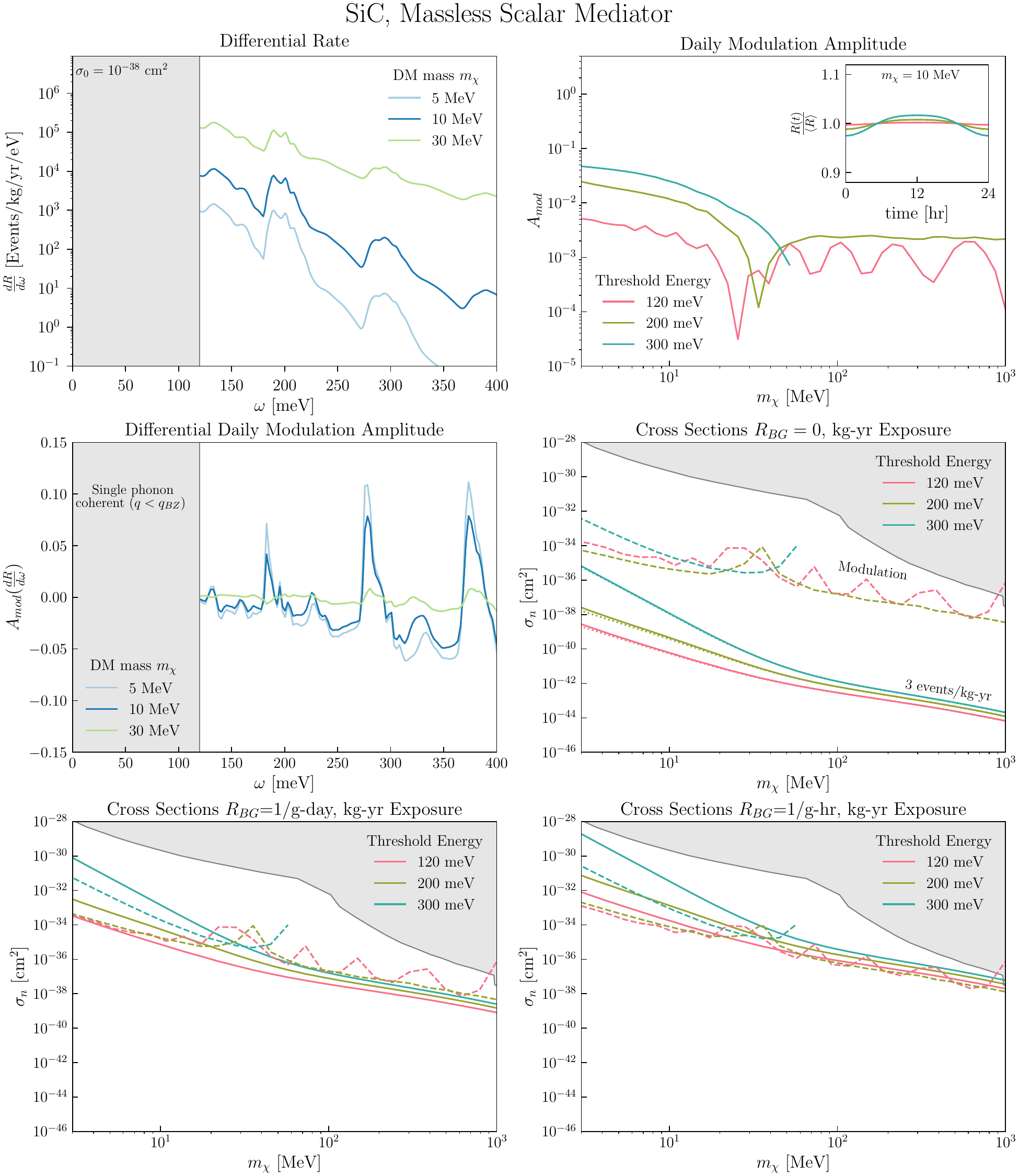}
    \caption{Results for SiC detector and massless scalar mediator.}
    \label{fig:sic_massless_scalar}
\end{figure*}

\begin{figure*}
    \centering
    \includegraphics[width=1\linewidth]{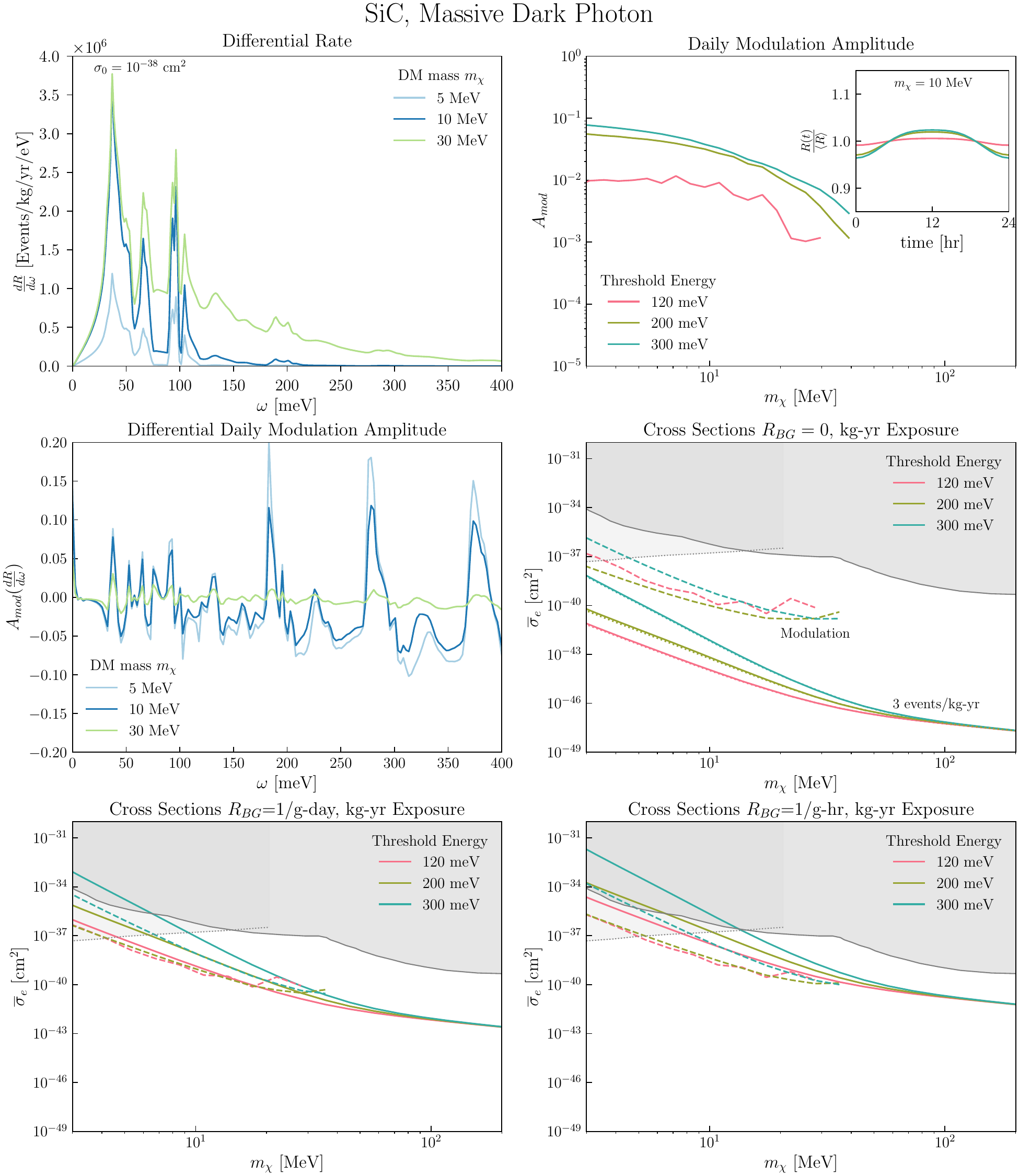}
    \caption{Results for SiC detector and massive dark photon mediator.}
    \label{fig:sic_massive_darkphoton}
\end{figure*}

\begin{figure*}
    \centering
    \includegraphics[width=1\linewidth]{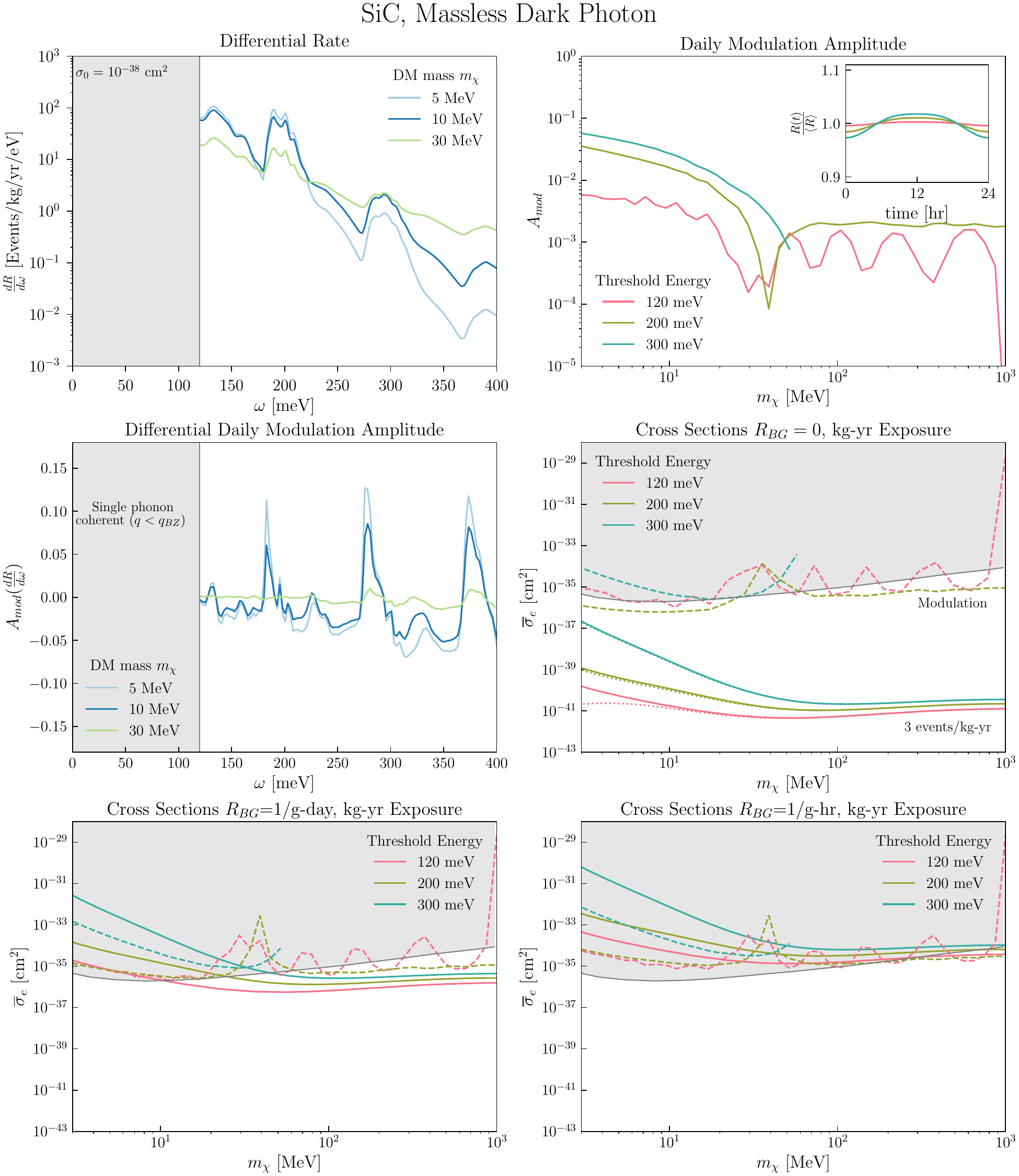}
    \caption{Results for SiC detector and massless dark photon mediator.}
    \label{fig:sic_massless_darkphoton}
\end{figure*}

\begin{figure*}
    \centering
    \includegraphics[width=1\linewidth]{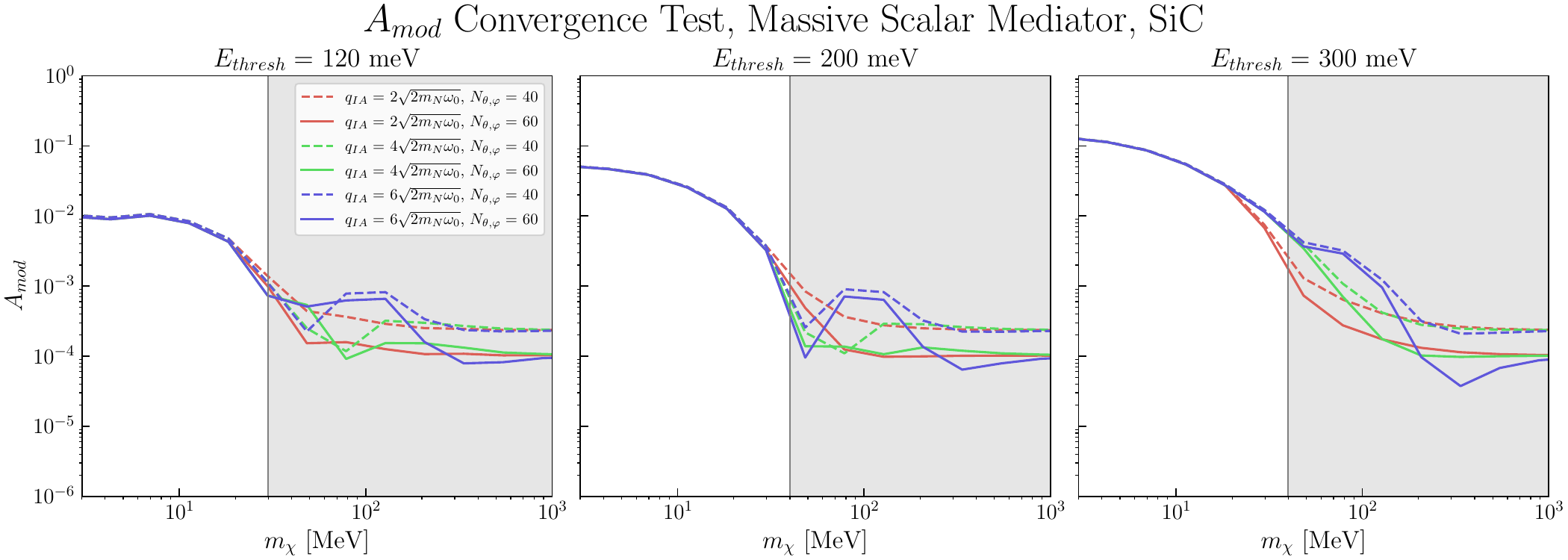}\\
    \vspace{1cm}
    \includegraphics[width=1\linewidth]{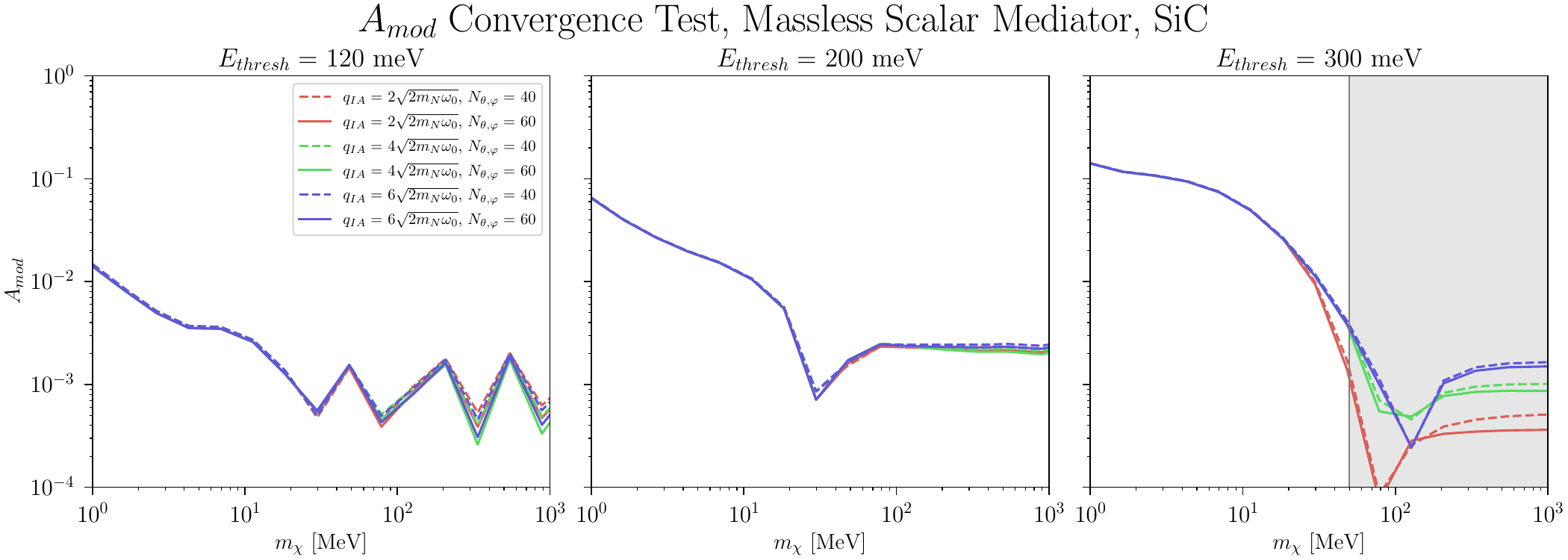}
    \caption{Convergence tests for $A_{mod}$ in SiC, for a massive scalar mediator (top) and a massless scalar mediator (bottom).}
    \label{fig:A_mod_convergence_test_sic}
\end{figure*}

First, in Fig.~\ref{fig:sic_pDoS} we show the components of the pDoS tensor. In Fig.~\ref{fig:sic_S_qw} we show the $n=1,2,3$ terms of the structure factor for SiC to offer insight into anisotropies at different $\omega$. Figs.~\ref{fig:sic_massive_scalar}, ~\ref{fig:sic_massless_scalar}, ~\ref{fig:sic_massive_darkphoton} and ~\ref{fig:sic_massless_darkphoton} show the same multi-panel plots of differential rate, daily modulation amplitude, differential daily modulation amplitude and cross sections as shown for Al$_2$O$_3$, again for various mediators. We find slightly less modulation across all mediators to sapphire. Fig.~\ref{fig:A_mod_convergence_test_sic} shows the same convergence tests discussed in App.~\ref{convergence_test_appendix} for SiC, with similar ranges of validity in $m_\chi$ as for sapphire.

\end{document}